\documentclass[journal]{IEEEtran}

\ifCLASSINFOpdf
\else
   \usepackage[dvips]{graphicx}
\fi
\usepackage{url}
\usepackage{graphicx}
\usepackage[linesnumbered,lined, noend,boxed,commentsnumbered]{algorithm2e}
\usepackage{cite}
\usepackage{amsmath,amssymb,amsfonts}
\usepackage{amsthm}
\usepackage{algorithmic}
\usepackage{graphicx}
\usepackage{textcomp}
\usepackage{mathrsfs}
\usepackage{mathtools, cuted}
\usepackage{tabularx}

\usepackage{mathtools}
\usepackage{stfloats}
\usepackage{mathtools, nccmath}
\usepackage[export]{adjustbox}
\usepackage[style=base]{caption}

\usepackage{subcaption}

\usepackage{color}
\usepackage{color}
\usepackage{tikz}
\usetikzlibrary{arrows.meta,
                chains,
                positioning,
                quotes,
                shapes.geometric}              
\makeatletter
\tikzset{FlowChart/.style={
suspend join/.code = {\def\tikz@after@path{}},
     base/.style = {draw, fill=##1,
                    minimum height=9mm, text width=34mm,
                    align=center,
                    on chain, join=by arr
                   },
startstop/.style = {base=red!30},
  process/.style = {base=orange!30, rounded corners},
 decision/.style = {diamond, aspect=1.3, inner xsep=0pt,
                    draw, fill=green!30, align=center,  
                    on chain, join=by arr},             
       io/.style = {base=blue!30, trapezium, trapezium stretches body,
                    trapezium left angle=70, trapezium right angle=110},
      arr/.style = {semithick,-Triangle},
every edge quotes/.style = {auto, font=\footnotesize}
       }   }
\makeatletter

\begin{document}

\title{VAFER: Signal Decomposition based Mutual Interference Suppression in FMCW Radars}

\author{Abhilash Gaur, Po-Hsuan Tseng,~\IEEEmembership{Member, IEEE}, Kai-Ten Feng,~\IEEEmembership{Senior Member, IEEE,} \\
and Seshan Srirangarajan,~\IEEEmembership{Member, IEEE}
}

\maketitle

\begin{abstract}
With increasing application of frequency-modulated continuous wave (FMCW) radars in autonomous vehicles, mutual interference among FMCW radars poses a serious threat. Through this paper, we present a novel approach to effectively and elegantly suppress mutual interference in FMCW radars. We first decompose the received signal into modes using variational mode decomposition (VMD) and perform time-frequency analysis using Fourier synchrosqueezed transform (FSST). The interference-suppressed signal is then reconstructed by applying a proposed energy-entropy-based thresholding operation on the time-frequency spectra of VMD modes. The effectiveness of proposed method is measured in terms of signal-to-interference plus noise ratio (SINR) and correlation coefficient for both simulated and experimental automotive radar data in the presence of FMCW interference. Compared to other existing literature, our proposed method demonstrates significant improvement in the output SINR by at least 14.07 dB for simulated data and 9.87 dB for experimental data.   
\end{abstract}

\begin{IEEEkeywords}
Interference mitigation, variational mode decomposition, frequency-modulated continuous wave (FMCW) radar.
\end{IEEEkeywords}

\IEEEpeerreviewmaketitle

\section{Introduction}

\IEEEPARstart{A}{utomotive} radars pave the way for advanced driver assistance systems (ADAS) development due to their robust operation under bad weather conditions and affordability. Frequency-modulated continuous wave (FMCW) radars are preferred in the automotive industry due to their ability to measure range, angle, and velocity~\cite{bilik_2019}. In addition to radars' attractive performance and reliability, recent developments in integrated radio frequency complementary metal-oxide-semiconductor (RF-CMOS) technology have enabled low-cost radar-on-chip systems in the $76-81$ GHz band~\cite{Luo_2013}. 

As automotive radars find application in lane-change assistance, ADAS, automatic emergency braking, etc., the number of such radars on vehicles are expected to increase in the near future. With the increasing density of automotive FMCW radars, mutual interference among these radars will be significantly severe~\cite{alland_2019,Bourdoux_2017}. During FMCW data processing, targets appear in the form of tones (beat signal) at the baseband, whereas external FMCW interference with a different chirp slope would appear as a chirp. Strong interference increases the noise floor, weakens the target mask, and reduces the probability of target detection.

In existing literature, several methods have been proposed to suppress the mutual interference among FMCW radars. In \cite{uysal_2020,khoury_2016,Aydogdu_2021}, different coordination methods are proposed for radar systems to avoid interference by modifying the radar parameters. A phase-coded FMCW system is presented in \cite{uysal_2020} to mitigate interference. In \cite{khoury_2016}, authors proposed a medium access control (MAC)-like approach which dynamically assigns radar parameters to multiple radars in the same area by communicating via a dedicated long-term evolution (LTE) link to a cloud-based system. A distributed networking protocol is proposed in \cite{Aydogdu_2021} to avoid mutual interference. These coordination methods avoid FMCW interference at the cost of an additional coordination unit to the already existing FMCW radar systems. 

Some researchers have also proposed various interference suppression methods by designing new transmit wave-forms and modifying the radar system architecture ~\cite{Bechter_2016,xu_2018,Irazoqui_2019,Hu_2019,Kunert_2012}. Authors in ~\cite{Bechter_2016} use the frequency hopping technique learned from bats to avoid mutual interference. In ~\cite{xu_2018}, orthogonal noise waveforms based on the phase retrieval method are transmitted to reduce the probability of interference. A spatial interference mitigation circuit (SIMC) is proposed in ~\cite{Irazoqui_2019} to alleviate interference before the signal reaches the receiver. In ~\cite{Hu_2019}, authors propose a multi-carrier frequency random transmission chirp sequence to avoid interference for time-division multiplexing (TDM) multi-input multi-output (MIMO) FMCW radars. These new wave-form designs and system level changes result in increased noise floor and affect the performance of radar's parameter estimation, including range, velocity, and angle. 

Other than the coordination-based methods, new wave-forms and system designs to suppress FMCW interference, several other signal processing-based algorithms are proposed in the literature~\cite{jin_2019,chen_2018,wang_2021,lee_2021,xu_2021,Neemat_2019,Uysal_2019,Liu_2020,Mun_2020,wang_2022_2}. 
These signal processing methods facilitate interference suppression without needing any change in the hardware system. In~\cite{jin_2019}, an adaptive noise canceller (ANC) is proposed to mitigate interference in the positive half of frequency spectrum by measuring the interference in the negative half of frequency spectrum. In~\cite{chen_2018}, a radio frequency interference (RFI) suppression algorithm based on orthogonal projection filtering is presented, in which the interference is estimated from the negative range bins and mitigated by projecting positive range bins onto the interference subspace. The performance of these filtering-based methods depends upon the availability of a proper reference input. In~\cite{wang_2021}, interference suppressed signal is reconstructed using a matrix pencil-based algorithm applied to the detected interference-contaminated signal regions. Meanwhile, reconstructed interference signal in~\cite{lee_2021} is subtracted from the interference corrupted signal using the proposed wavelet transform-based algorithm. A signal separation-based algorithm using tunable Q-factor wavelet transform is proposed in~\cite{xu_2021}. Recently, the framework presented in~\cite{wang_2022} mitigates interference in FMCW radars using sparse and low-rank Hankel matrix decomposition. 

In this paper, we propose a unique mutual interference suppression algorithm for FMCW radars using VAriational mode decomposition (VMD), Fourier-based synchrosqueezing
transform (FSST), and the Energy-entRopy thresholding-based reconstruction, named as VAFER method. In the interference-contaminated signal, the targets present as single-tone beat frequencies while interference appears as a chirp. We decompose the interference-contaminated received signals by using VMD \cite{dragomi_2014} in different modes, containing information about the target's beat frequencies and interference. We emphasize  using VMD as a decomposition tool due to the ability of VMD to map these single-tone frequencies into individual modes. We further adopt FSST \cite{thakur_2011} for time-frequency analysis to distinguish among different modes. The property of FSST to focus on sinusoid-like signals allow us to  represent the modes corresponding to target's beat frequencies as individual lines parallel to the time axis; while the interference reveals as noises. However, the combination of VMD and FSST still cannot identify the various modes corresponding to beat frequencies. Therefore, we propose an energy-entropy-based threshold utilizing Wiener entropy, which works as a measure of the spectral variation for the spectra (i.e., spikiness of spectra) produced by FSST of the modes. Finally, the remaining modes which fall within the proposed energy-entropy threshold are combined together to reconstruct the interference-suppressed signal. 
The main contributions of this paper are listed as follows:
\begin{itemize}
    \item We present an analytical framework proposing the first-ever interference suppression technique for FMCW radars utilizing VMD to decompose interference-contaminated signals into different modes and then adopting FSST on these modes to perform time-frequency analysis.
	\item We proposed an energy-entropy-based threshold to discard those VMD modes containing interference, and consequently reestablish the interference-suppressed signals.
	\item The interference suppression ability of proposed VAFER algorithm is demonstrated via both simulations and experimental data. Comparing with recently proposed ANC ~\cite{jin_2019}, sparse and low-rank Hankel matrix decomposition method (SPARKLE)~\cite{wang_2022}, and the short-time Fourier transform (STFT)-based version of VAFER method, the VAFER scheme achieves the best target detection performance and improves the output signal-to-interference plus noise ratio (SINR) with at least $9.87$ dB.
\end{itemize}

This paper is organized as follows. Section II shows the signal model and problem formulation. In Section III, the proposed VAFER method for interference mitigation is described. Simulation and experimental results demonstrating the effectiveness of VAFER algorithm are shown in Section VI, followed by conclusions in Section V.

\section{Signal Model and Problem Formulation}
\begin{figure}
    \centering
    \includegraphics[width=1\linewidth,center]{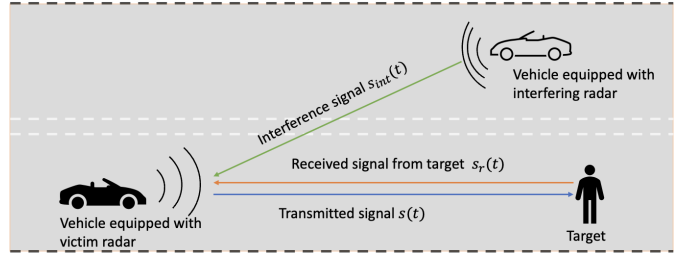}
  \caption{Network scenario for interfering FMCW radars.}
    \label{fig:int_scene}
\end{figure}

It is considered that $Q$ point targets and $M$ interfering radars exist in the observation scene. As shown in Fig. \ref{fig:int_scene}, the normalized transmitted chirp signal $s(t)$ at time $t$ of the victim radar is 
\begin{equation}
s(t)=\exp\left[{j2\pi\left(f_ot+\frac{1}{2}\mu t^2\right)}\right],
\end{equation}
where $f_o$ is the starting frequency, $\mu=\frac{B}{T}$ represents the chirp slope with $B$ as the sweep bandwidth of a single ramp and $T$ as time duration for $t\in$ [0, $T$]. On the other hand, the reflected signal received at the victim radar from the $Q$ targets is the superposition of delayed transmitted signal $s(t)$ which can be expressed as
\begin{equation}
s_r(t)=\sum_{q=1}^{Q}\alpha_q\exp\left[{j2\pi\left(f_o(t-\tau_q)+\frac{1}{2}\mu (t-\tau_q)^2\right)}\right],
\label{eqn:s}
\end{equation}
where $\tau_q$ is the propagation delay and $\alpha_q$ denotes complex amplitude for the $q^{th}$ target. 

Meanwhile, let $s_{int}(t)$ be the interference signal from interfering radars affecting the reflected signal $s_r(t)$ received from targets. Considering $m^{th}$ FMCW interfere radar having chirp slope $\beta_m$ and starting frequency $f_{m}$ with time delay $\tau'_m$ with respect to transmitted chirp, the total interference signals from $M$ interfering radars observed at the receiver of victim radar can be modeled as
\begin{equation}
s_{int}(t)=\sum_{m=1}^{M}\exp\left[{j2\pi\left(f_{m}(t-\tau'_m)+\frac{1}{2}\beta_m(t-\tau'_m)^2\right)}\right].
\label{eqn:sint}
\end{equation}
A typical interference scenario is depicted in Fig.~\ref{fig:int_scene}, where  the received signal observed by the receiving antenna of victim radar is the combination of target reflections $s_r(t)$ and interference signal $s_{int}(t)$, i.e., $r(t)= s_r(t)+s_{int}(t)$. 
The received signal at the antenna of victim radar $r(t)$ is first de-chirped by mixing with a conjugate copy of transmitted signal $s(t)$ and then passed through an analog low-pass filter (LPF) having impulse response $h(t)$ to obtain beat signal $r_{T}(t)$ as   

\begin{equation}
    r_{T}(t)=[(s_r(t)+s_{int}(t))\cdot s^*(t)]\otimes h(t).
    \label{eqn:st}
\end{equation}
Note that the main purpose of passing through an LPF is to bound the received interference signal's instantaneous frequency. Based on (\ref{eqn:s}), (\ref{eqn:sint}), and (\ref{eqn:st}), the total received signal $r_{T}(t)$ can be obtained after de-chirping for $Q$ targets and $M$ interferers and then passing through LPF as 
    \begin{align}
    r_T&(t)=  \sum_{q=1}^{Q}\alpha_q\exp\left[{j2\pi\left(-f_o\tau_q-\mu t\tau_q+\frac{\mu \tau_q^2}{2}\right)}\right]\otimes h(t)\nonumber\\
    &+\sum_{m=1}^{M}\exp\left[{j2\pi\left((f_{m}-f_o)t+\frac{1}{2}(\beta_m-\mu)t^2 \right)}\right]\cdot \nonumber\\  &\exp\left[{j2\pi\left(-f_{m}\tau_m'-\beta_m  t\tau_m'+\frac{\beta_m}{2} t\tau_m'^2\right)}\right]\otimes h(t)+\eta(t).
    \label{eqn:tot_sig2}
    \end{align}
Notice that the total received signal in \eqref{eqn:tot_sig2} consists of three terms as follows: (a) the first term  represents the target echo appearing as a tone at a particular beat frequency, (b) the second term corresponds to the interference appearing as a chirp in the baseband signal, and (c) the third term $\eta (t)$ represents the additive white Gaussian noise. \eqref{eqn:tot_sig2} also shows the dependency of interference on the starting frequencies and chirp slopes of the victim and interfering radar. Our objective is to develop a method to alleviate the effect of interference terms as shown in \eqref{eqn:tot_sig2}, which will be elaborated in the following section.

\section{Proposed VAFER Scheme}
Our main objective is to effectively detect the existence of point target based on an FMCW radar by suppressing surrounding interference from interfering radars. In the following subsections, we will provide detailed explanation of our proposed VAFER algorithm, which consists of mode decomposition of interference-contaminated signal using VMD, time-frequency analysis with FSST, and energy-entropy thresholding for signal reconstruction.

\subsection{Variational Mode Decomposition (VMD)}

To decompose the interference-contaminated noisy signal, we propose using an adaptive signal decomposition method, i.e., VMD, that decomposes a signal into $K$ narrow-band variational mode functions (VMFs). Since the beat signal corresponding to each target is a single-tone frequency, VMD is considered feasible in separating the frequencies for their respective radar targets in its VMFs. 
Let $\{u_k \} \coloneqq \{ u_1(t),...,u_K(t) \}$ and $\{\omega_k \} \coloneqq  \{\omega_1,...,\omega_K \}$ denote collections of $K$ VMFs and their center frequencies respectively. For each VMF ${u_k }$, VMD uses Hilbert transform to compute the associated analytic signal. Then, the estimated center frequency is used to shift the frequency spectrum of VMFs to the baseband, followed by applying Gaussian smoothness for bandwidth estimation. Thus, VMD decomposes a signal into its narrow-band components by considering bandwidth as sparsity prior in the spectral domain, where bandwidth is estimated by the squared $L_2$ norm of gradient. The resulting constrained variational problem is stated as 
\begin{equation}
\begin{aligned}
&\min \limits _{\{\omega_{k} \},\{u _{k} \}} \left \{{{\sum \limits _{k=1}^{K} {\left \|{ {\partial _{t} \left [{ {\left ({{\delta (t)+\frac {j}{\pi t}} }\right)\ast u_{k} (t)} }\right]e^{-j\omega _{k} t}} }\right \|_{2}^{2}}} }\right \} \\&s.t.\quad \sum \limits _{k=1}^{K} {u_{k}(t)} =r_T(t),
\end{aligned}
\label{cvp}
\end{equation}
where $\partial_{t}$ denotes partial derivative with respect to $t$, $\delta (\cdot)$ is the unit impulse function, and $*$ indicates convolution operation. In order to obtain the optimal solution to the constrained variational problem in (\ref{cvp}), it is transformed into a non-constrained optimization problem using a quadratic penalty factor $\alpha$  
and the Lagrangian multiplier $\lambda$ following \cite{dragomi_2014} as
%
\begin{align}
L&\left ({{\{\omega_{k} \},\{u_{k} \},\lambda } }\right) =\nonumber\\
&\alpha \sum \limits _{k=1}^{K} {\left \|{ {\partial _{t} \left [{ {\left ({{\delta (t)+\frac {j}{\pi t}} }\right)\ast u_{k} (t)} }\right]e^{-j\omega _{k} t}} }\right \|_{2}^{2}} \nonumber\\
&+\,\left \|{ {r_T(t)-\sum \limits _{k=1}^{K} {u_{k} (t)}} }\right \|_{2}^{2} +\left \langle{ {\lambda (t),r_T(t)-\sum \limits _{k=1}^{K} {u_{k} (t)}} }\right \rangle,
\label{eq:VMD2}
\end{align}
where the operator $\langle \cdot,\cdot \rangle $ denotes inner product. The optimization of \eqref{eq:VMD2} takes place in the Fourier domain, and is solved via the alternate direction method of multipliers (ADMM) for iterative updates of VMFs in frequency domain and frequencies, i.e., $\hat {u}_{k}^{n+1} (\omega)$ and $\omega_k^{n+1}$, by adopting the convergence criterion and convergence tolerance $\xi$ in \cite{dragomi_2014} as
\begin{align}
\hat {u}_{k}^{n+1} (\omega)=\frac {\hat {r}_T(\omega)-\sum \limits_{i = 1}^{k-1} \hat {u}_{i}^{n+1} (\omega) -\sum \limits_{i = k+1}^K \hat {u}_{i}^{n} (\omega) +\hat {\lambda }(\omega)/2}{1+2\alpha (\omega -\omega _{k}^n)^{2}} \label{eq:u_k}
\end{align}
\begin{align}
\omega _{k}^{n+1} =\frac {\int _{0}^\infty {\omega \left |{ {\hat {u}_{k}^{n+1} (\omega)} }\right |^{2}d\omega }}{\int _{0}^\infty {\left |{ {\hat {u}_{k}^{n+1} (\omega)} }\right |^{2}d\omega }}  \label{eq:omega_k}
\end{align}
where $(\cdot)^{n+1}$ denotes the $(n+1)^{th}$ iteration, $\hat {r}_T(\omega)$ and $\hat{\lambda }(\omega)$ represent the Fourier transform of $r_T(t)$ and $\lambda(t)$, respectively, with $\omega$ indicating the frequency variable. The most recent VMFs and frequencies are utilized for the following updates. 

With the designed VMD method, the signal of interest and interference signals are distributed in different VMFs. Because beat frequencies corresponding to targets are present as single-tone frequencies, target reflections should be contained in a few VMFs as sinusoidal variations of constant frequency whereas the interference appears as a chirp-like variation, which will be numerically validated in the performance evaluation section.


\subsection{Time-Frequency Analysis with FSST}
In this step, we propose to perform a time-frequency analysis of $K$ VMFs using  FSST. Notice that we prefer FSST over the conventional time-frequency analysis tool STFT because it highly focuses on sinusoid-like target signals instead of focusing on the interference patterns~\cite{thakur_2011}. 
FSST sparsely represents VMFs that contain information about target reflections as a single frequency in the time-frequency representation. 
First of all, we calculate the STFT of $k^{th}$ VMF as
\begin{equation} V_{u_{k}}^{g}(t,\omega)=\int_{\mathbb{R}}u_{k}(\tau)g^{\ast}(\tau-t)e^{-i\omega\tau}d\tau, \label{eqn:fsst_1}\end{equation}
where $\mathbb{R}$ is the set of real numbers 
and $g^*(t)$ indicates the complex conjugate of $g(t)$ denoting the sliding window function \cite{thakur_2011}.
The FSST is obtained from frequency domain reassignment of the STFT to its spectrogram centroid $\hat{\omega}_{u_{k}}(t, \omega)$ over each time instant $t$ \cite{auger_2013}. The centroid of STFT spectrogram is
\begin{equation}\hat{\omega}_{u_{k}}(t, \omega)=\omega-\Im\left\{\frac{V_{u_{k}}^{g^{\prime}}(t,\omega)}{V_{u_{k}}^{g}(t,\omega)}\right\}, \label{eqn:fsst_2}\end{equation}
where $V_{u_{k}}^{g^{\prime}}(t,\omega)$ denotes the STFT of $k^{th}$ VMF obtained with the derivative of sliding window function $g(t)$, and $\Im$ indicates the imaginary part of the input \cite{oberlin_2014}. 
Therefore, based on (\ref{eqn:fsst_2}), the FSST of $k^{th}$ VMF obtained after the reassignment of STFT is
\begin{equation} U_{k}(t, \omega)=\frac{\int_{\mathbb{R}}V_{u_{k}}^{g}(t,z)e^{i\omega t}\delta(\omega-\hat{\omega}_{u_{k}}(t,z))dz}{2\pi g^{\ast}(0)}. \label{eqn:fsst_3} \end{equation}
%
Note that the modes acquired by VMD are transformed into time-frequency spectra by applying FSST scheme in (\ref{eqn:fsst_3}). The property of FSST to emphasize sinusoid-like signals allow us to characterize single-tone beat frequencies corresponding to targets, i.e., straight frequency line parallel to the time axis; whereas interference is illustrated as a noisy blob in the spectra. We will evaluate the effectiveness of proposed FSST method in performance evaluation.


\subsection{Energy-Entropy Thresholding and Reconstruction}

After obtaining the time-frequency spectrum for $K$ VMFs, we propose to adopt the Wiener Entropy to distinguish whether a particular mode contains information about target reflections or interference based on their specific characteristics. Note that Wiener entropy provides a measure of the {\it spikiness} of a vector. If the sample values in a vector do not vary much, the Wiener entropy reaches 1; while the Wiener entropy will approach 0~ with larger variation in sample values \cite{Dubnov_2004}.
Based on (\ref{eqn:fsst_3}), we compute the Wiener entropy $W(U_k(t, \omega))$ of the $k$-th time-frequency representation of VMFs as
\begin{equation}
W(U_k(t, \omega))={\frac{\left[\prod\limits_{\upsilon=1}^{N}U_k(t,\omega)\right]^{1/N}} {{\frac{1}{N} \sum\limits_{\upsilon=1}^{N}U_k(t,\omega)}}}.
\label{W}
\end{equation}
For notational simplicity, we omit the index $\upsilon$ used for each time sample of $U_k(t,\omega)$. A total of $N$ time samples of $U_k(t,\omega)$ are utilized in the calculation of the Wiener entropy.


We propose an energy-entropy-based threshold to identify and suppress interference-contaminated VMFs. Based on (\ref{W}), the proposed threshold $T_{\beta}$ is calculated as
\begin{equation}
    T_{\beta}= \frac{\sum_{{k=1}}^{K} E_k \cdot W(U_k(t, \omega))}{\sum_{{k=1}}^{K} E_k},
    \label{eq:threshold}
\end{equation}
where $E_k$ denotes the energy of $k^{th}$ VMF computed as
\begin{equation}
    E_k= \int_{-\infty}^{\infty}\mid u_k(t)\mid^2 dt.
    \label{eq:energy}
\end{equation}
%
Finally, we reconstruct the interference-suppressed signal by choosing elements of the family of VMFs as
\begin{equation}
    \mathbb{V} = \big\{U_k(t,\omega): W(U_k(t, \omega))\geq T_{\beta},\ \forall k \in \mathbb{K} \big\},
    \label{V}
\end{equation}
where $\mathbb{K}$ is the index set containing the maximum number as $K$. (\ref{V}) denotes the set $\mathbb{V}$ of $U_k(t, \omega)$ whose Wiener entropy passes the threshold $T_{\beta}$.
As a result, the interference-suppressed signal is reconstructed by taking the inverse FSST of these VMFs as
\begin{equation}
   \hat{r}_T(t)=IFSST\left( \sum_{ \mathbb{V}}U_k(t,\omega) \right).
\end{equation}
The complete algorithm of proposed VAFER scheme is shown in Algorithm~\ref{alg:one}. 

\RestyleAlgo{ruled}

\begin{algorithm}[tbh]
\SetKwInOut{Input}{Input}
\SetKwInOut{Output}{Output}  \SetKw{Kw}{thetext} 
\AlgoDisplayBlockMarkers\SetAlgoBlockMarkers{begin}{end}%
\SetAlgoNoEnd
\caption{Proposed VAFER Algorithm}\label{alg:one}
\Input {$r_T(t)$: the interference contaminated signal }
Set $K$, $\alpha$ \\
Initialize $\mu_k^1$, $\omega_k^1$, $\lambda$, $n=0$ \\
Repeat $n \leftarrow n+1$ \\
\For{$k\in 1:K$}
{
update $u_k$ \\
$\hat {u}_{k}^{n+1} (\omega)\leftarrow \frac {\hat {r}_T(\omega)-\sum \limits_{i = 1}^{k-1} \hat {u}_{i}^{n+1} (\omega) -\sum \limits_{i = k+1}^K \hat {u}_{i}^{n} (\omega) +\hat {\lambda }(\omega)/2}{1+2\alpha (\omega -\omega _{k}^n)^{2}} $ \\
update $\omega_k$ \\
$\omega _{k}^{n+1} \leftarrow \frac {\int _{0}^\infty {\omega \left |{ {\hat {u}_{k}^{n+1} (\omega)} }\right |^{2}d\omega }}{\int _{0}^\infty {\left |{ {\hat {u}_{k}^{n+1} (\omega)} }\right |^{2}d\omega }}$ for all $\omega \geq 0$ \\
update $\lambda$ \\
$\hat{\lambda}^{n+1}(\omega)\leftarrow\hat{\lambda}^{n}(\omega)+\tau\left(\hat{r}_T(\omega)-\sum_{k}\hat{u}^{n+1}(\omega)\right)$ \\
repeat until $\sum_{k}\Vert\hat{\mu}_{k}^{n+1}-\hat{\mu}_{k}^{n}\Vert_{2}^{2}/\Vert\hat{\mu}_{k}^{n+1}\Vert_{2}^{2} < \xi$ 
}
Compute time-frequency spectra $U_k(t,\omega)$ using ~\eqref{eqn:fsst_1}, ~\eqref{eqn:fsst_2} and ~\eqref{eqn:fsst_3}
\\ Calculate wiener entropy $W(U_k(t, \omega))$ \\
\qquad $W(U_k(t, \omega))={\frac{\left[\prod\limits_{\upsilon=1}^{N}U_k(t,\omega)\right]^{1/N}} {{\frac{1}{N} \sum\limits_{\upsilon=1}^{N}U_k(t,\omega)}}}$ \\

Calculate energy-entropy threshold $ T_{\beta}$
\\
 \qquad $T_{\beta}= \frac{\sum_{{k=1}}^{K} E_k \cdot W(U_k(t, \omega))}{\sum_{{k=1}}^{K} E_k}$ \\

\For{$k\in 1:K$} {
 \lIf{$W \left(U_k(t,\omega)\right)\geq T_{\beta}$ \\}  {$\mathbb{V} \leftarrow \big\{U_k(t,\omega) \big\}$
}\lElse{
    discard VMF
}
}
Signal reconstruction via inverse FSST \\
\qquad $\hat{r}_T(t)=IFSST\left( \sum_{ \mathbb{V}}U_k(t,\omega) \right)$ \\
\Output {$\hat{r}_T(t)$: reconstructed  interference suppressed signal}
\end{algorithm}

\begin{figure*}[!ht]
\begin{subfigure}{.46\textwidth}
\includegraphics[width=01\linewidth,center]{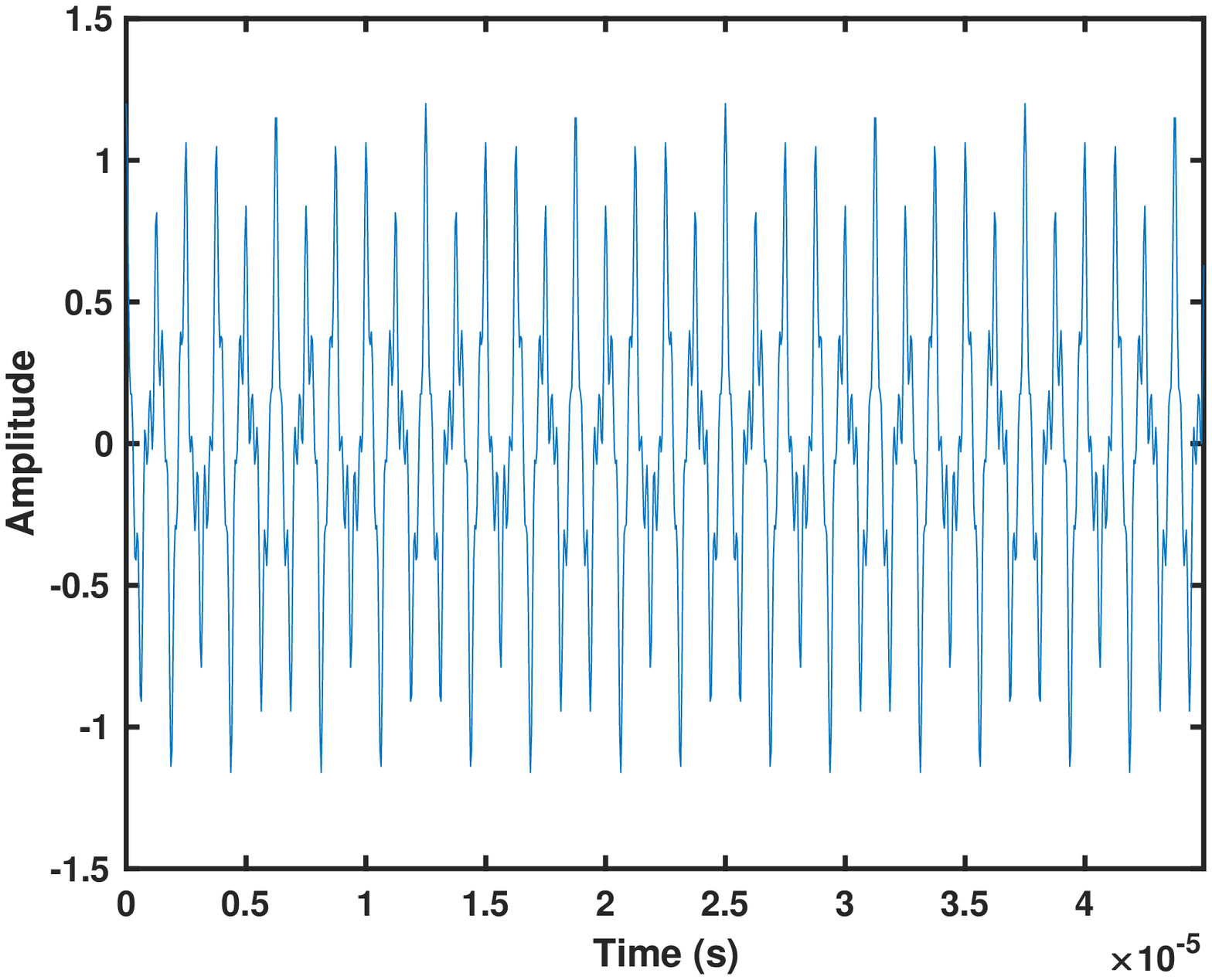}
  \caption{Beat signal of targets}
  \label{fig:sfig3}
\end{subfigure}%
\begin{subfigure}{.46\textwidth}
\includegraphics[width=01\linewidth,center]{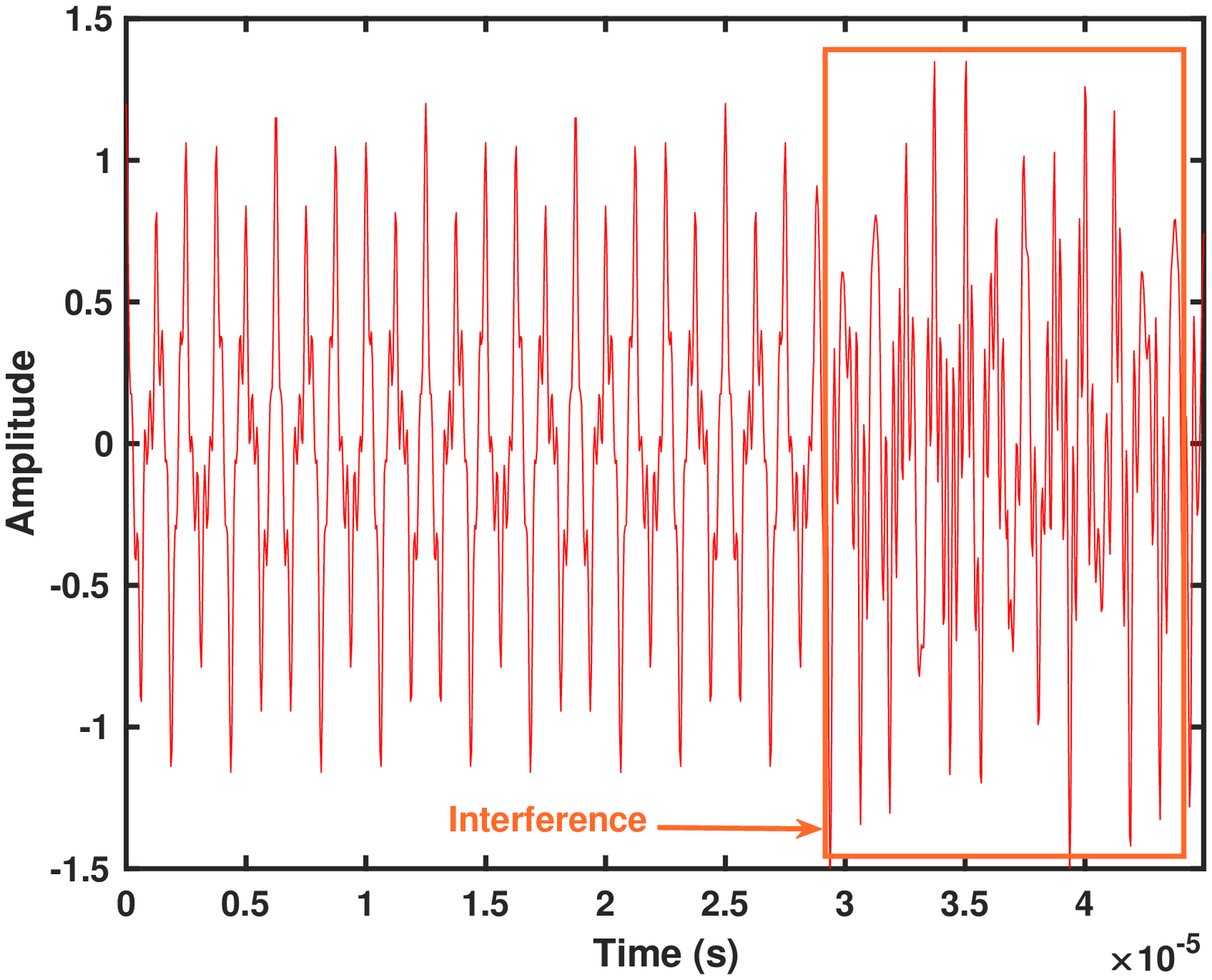}
  \caption{Beat signal of targets corrupted by interference}
  \label{fig:sfig4}
\end{subfigure}
\begin{subfigure}{.46\textwidth}
\includegraphics[width=01\linewidth,center]{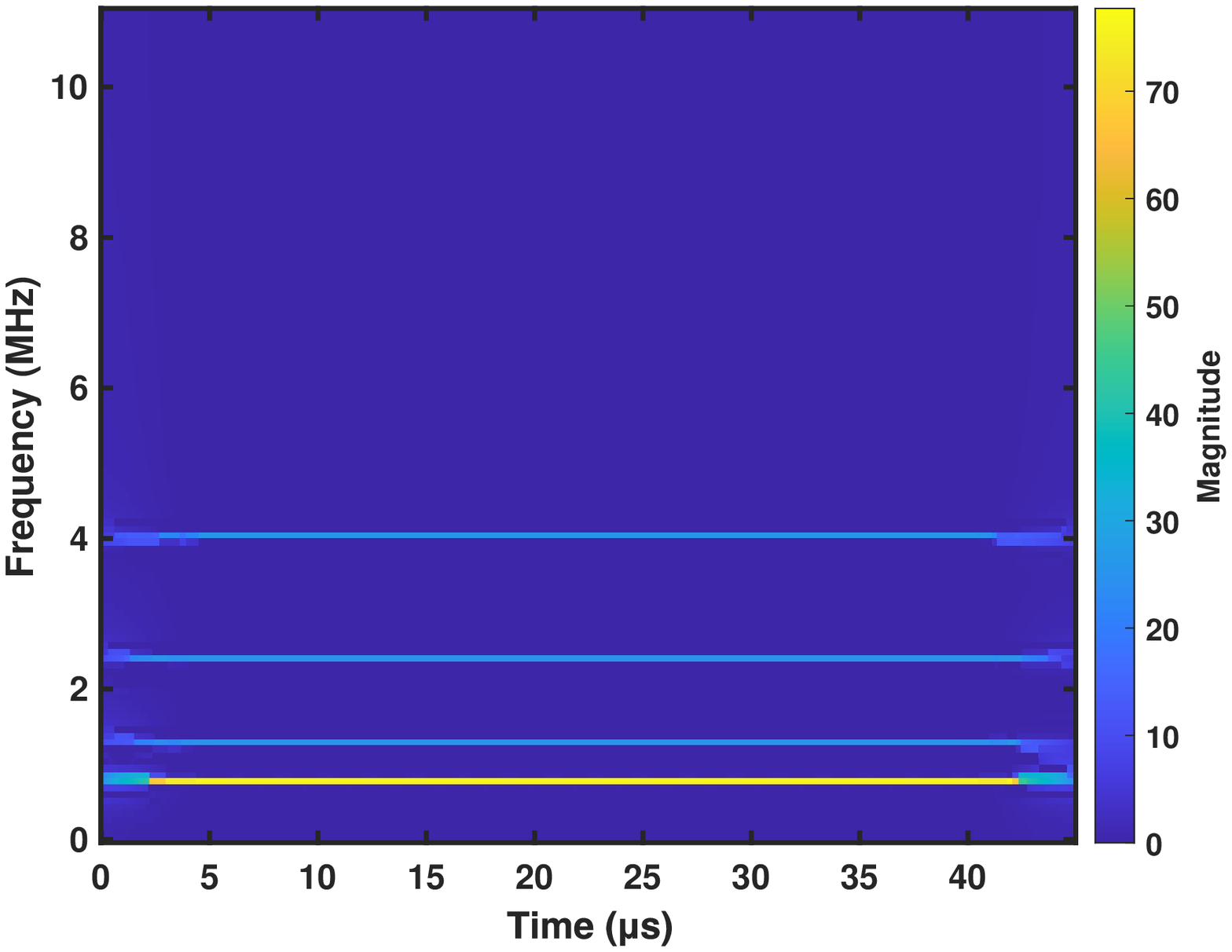}
  \caption{t-f spectra of pure beat signal}
  \label{fig:sfig_3}
\end{subfigure}%
\begin{subfigure}{.46\textwidth} \includegraphics[width=01\linewidth,center]{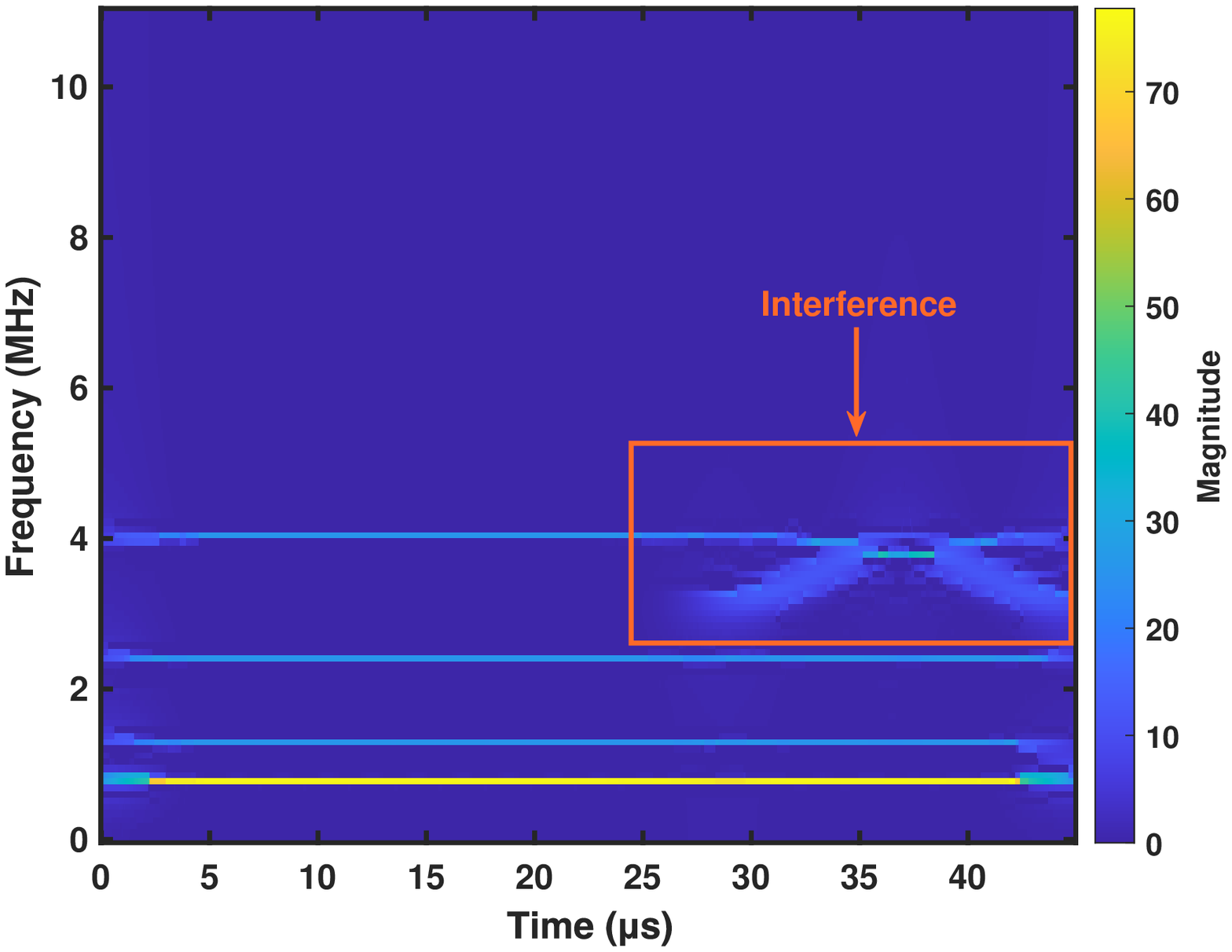}
  \caption{t-f spectra of interference-corrupted beat signal}
  \label{fig:sfig_4}
\end{subfigure}
\begin{subfigure}{.46\textwidth}
  \includegraphics[width=01\linewidth,center]{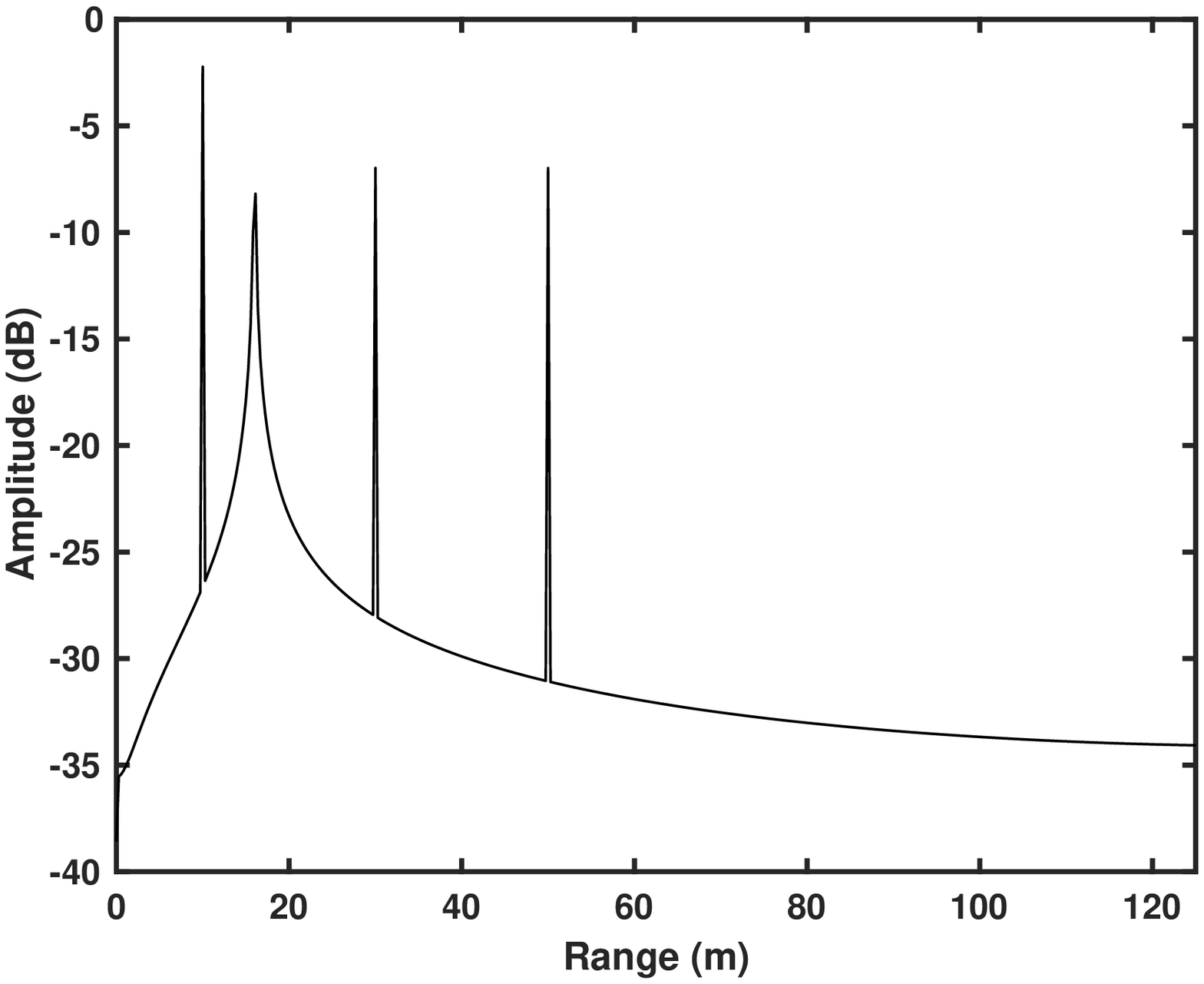}
  \caption{Range DFT of target only signal}
  \label{fig:s_fig3}
\end{subfigure}%
\begin{subfigure}{.46\textwidth}
  \includegraphics[width=01\linewidth,center]{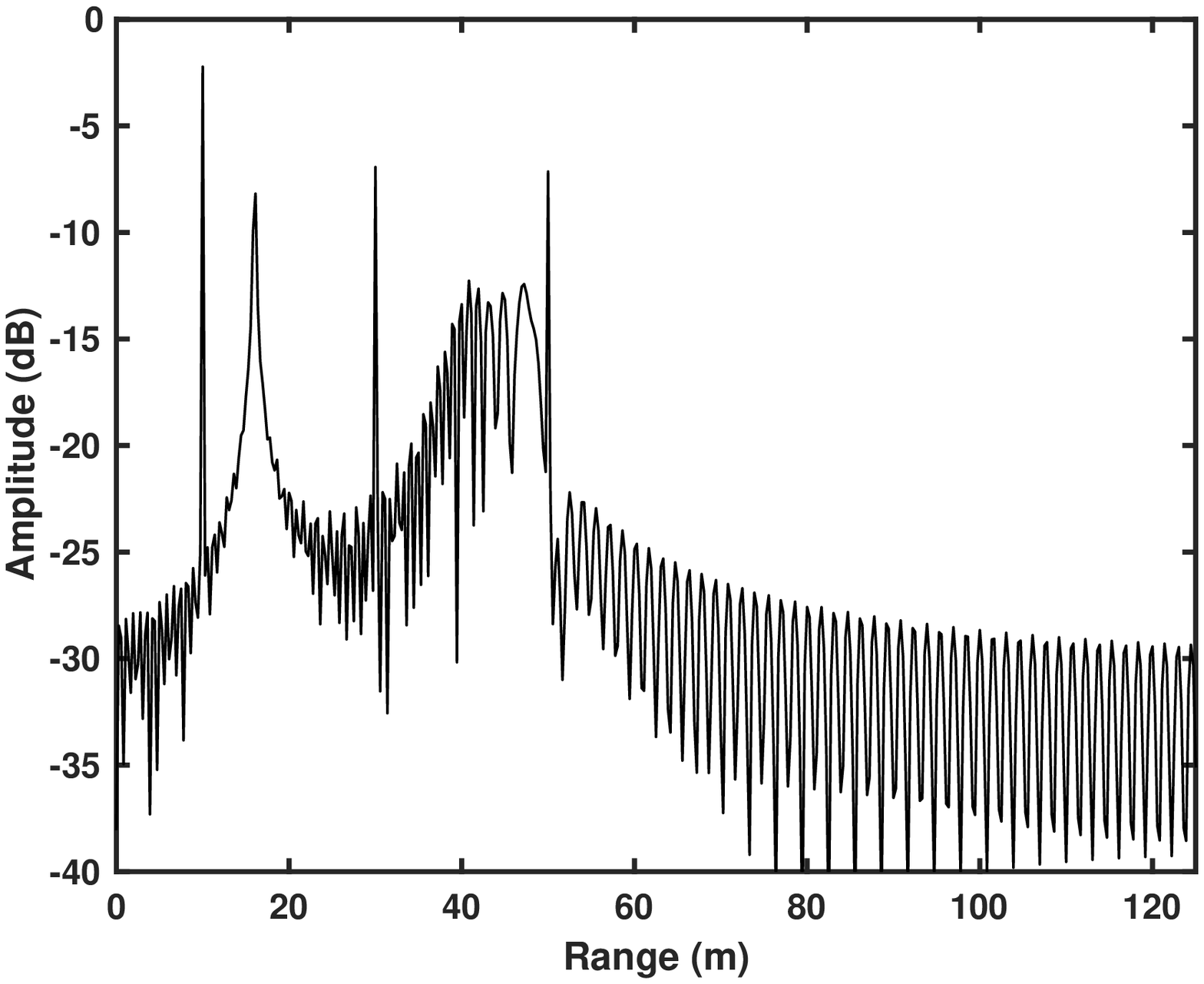}
  \caption{Range DFT of interference-corrupted beat signal}
  \label{fig:s_fig4}
\end{subfigure}%
\caption{Time domain, time-frequency domain representation, and range DFT of beat signal without and with interference.}
\label{fig:fig1}
\end{figure*}

\section{Performance Evaluation}
In this section, we present both simulation and experimental results to validate the effectiveness of our proposed interference suppression method for mutual interference between FMCW radars. The results of proposed VAFER method are compared with two state-of-the-art methods, i.e., the SPARKLE method \cite{wang_2021}, the adaptive interference mitigation method called ANC~\cite{jin_2019}, and a variant of proposed VAFER method where we use STFT instead of FSST, denoted as VAFER-STFT. The simulations and processing of experimental data have been conducted in MATLAB, and experimental data has been taken from~\cite{jin_2019}. 

\subsection{Performance Metrics}
We evaluate and compare the performance of our proposed VAFER scheme with other methods quantitatively in terms of SINR and correlation coefficient $\rho$. The SINR of interference-contaminated received signal is defined as
\begin{equation}
    {SINR_I} = 20\log _{10} \frac{\Vert  s_r\Vert _2}{\Vert s_{int} + \mathbf {\eta} \Vert _2},
\end{equation}
where $s_r$ is the reference signal from (\ref{eqn:s}), $ s_{int}$ is the interference signal from (\ref{eqn:sint}), and $\eta$ represents noise.
%
The SINR of reconstructed signal after interference suppression is defined as
\begin{equation}
    {SINR_O} = 20\log _{10} \frac{\Vert  s_r \Vert _2}{\Vert  s_r - \hat{r}_T \Vert _2},
    \label{eqn:op_sinr}
\end{equation}
where $\hat{r}_T$ is the reconstructed interference-suppressed signal from (\ref{eqn:st}). In  \eqref{eqn:op_sinr}, the denominator $\Vert s_r - \hat{r}_T \Vert _2$ represents the error between interference-free signal and the reconstructed signal. Hence, a lower value of error leads to higher $SINR_O$, characterizing better interference suppression.
Furthermore, the correlation coefficient between the interference-free signal and the reconstructed signal is defined as
\begin{equation} \rho = \frac{ {s_r}^H  \hat{r}_T}{\left\Vert s_r\right\Vert _2 \cdot \left\Vert { \hat{r}_T}\right\Vert _2}, 
\label{rho}
\end{equation}
where the correlation coefficient $\rho$ ranges from $[0,1]$ and $(\cdot)^H$ denotes conjugate transpose. Higher $\rho$ value specifies a higher correlation between the reference signal $s_r$ and reconstructed signal $\hat{r}_T$.
%
\subsection{Simulation Results} 
\subsubsection{Simulation Setup} 
\label{sec:simsetup}

\begin{table}[ht]
\centering
\caption{\\Simulation Parameters
}
\begin{tabular}[t]{lcc}
\hline
Parameter & Value  \\
 \hline
Carrier Frequency $f_o$ & 77 GHz   \\

Bandwidth $B$ & 540 MHz   \\

Chirp Duration $T$ & 45 $\mu$sec   \\

Sampling Frequency $f_s$ & 22 MHz \\

Maximum Beat Frequency $f_{b,max}$ & 10 MHz \\

\hline
\end{tabular}
\label{table:1}
\end{table}

In simulations, we have considered $Q=4$ point targets in the field of view (FOV) of victim radars at the ranges $R=[10, 16, 30, 50]$ meters. The parameters of radar used in simulations are shown in Table \ref{table:1}. Note that the maximum beat frequency $f_{b,max}$ is the cutoff frequency adopted in LPF. The reflected signal received at the victim radar is interfered by $M=2$ interfering radars transmitting chirp signals with starting frequency $f_m=77$ GHz for $m=1$ and $2$ and chirp slopes $\beta_1=1.5\mu$ and $\beta_2=2\mu$ with $\mu$ as the slope of victim radar. 

Fig.~\ref{fig:fig1} shows the beat signals of four point targets without (left-column plots) and with (right-column plots) interference along with their corresponding time-frequency (t-f) domain spectra and range discrete Fourier transform (DFT). Fig.~\ref{fig:fig1}(b) illustrates that $36\%$ of the entire signal time sequence is specified to be contaminated with the interference. The four point targets appear as horizontal lines each representing single-tone frequency as shown in Figs.~\ref{fig:fig1}(c) and \ref{fig:fig1}(d) in the time-frequency spectra. Note that the target with the lowest frequency possesses the largest signal magnitudes illustrated with yellow color. Furthermore, Figs.~\ref{fig:fig1}(e) and \ref{fig:fig1}(f) show the four peaks in range DFT of the received radar data indicating those four point targets. Notice that the interference has a chirp-like nature, as shown in Fig. \ref{fig:fig1}(e), where the interference-contaminated received signal has $SINR_I$ of $9.1814$ dB in simulations.


\begin{figure}[!ht]
\centering
\begin{subfigure}{.5\textwidth}
  \includegraphics[width=1\linewidth,center]{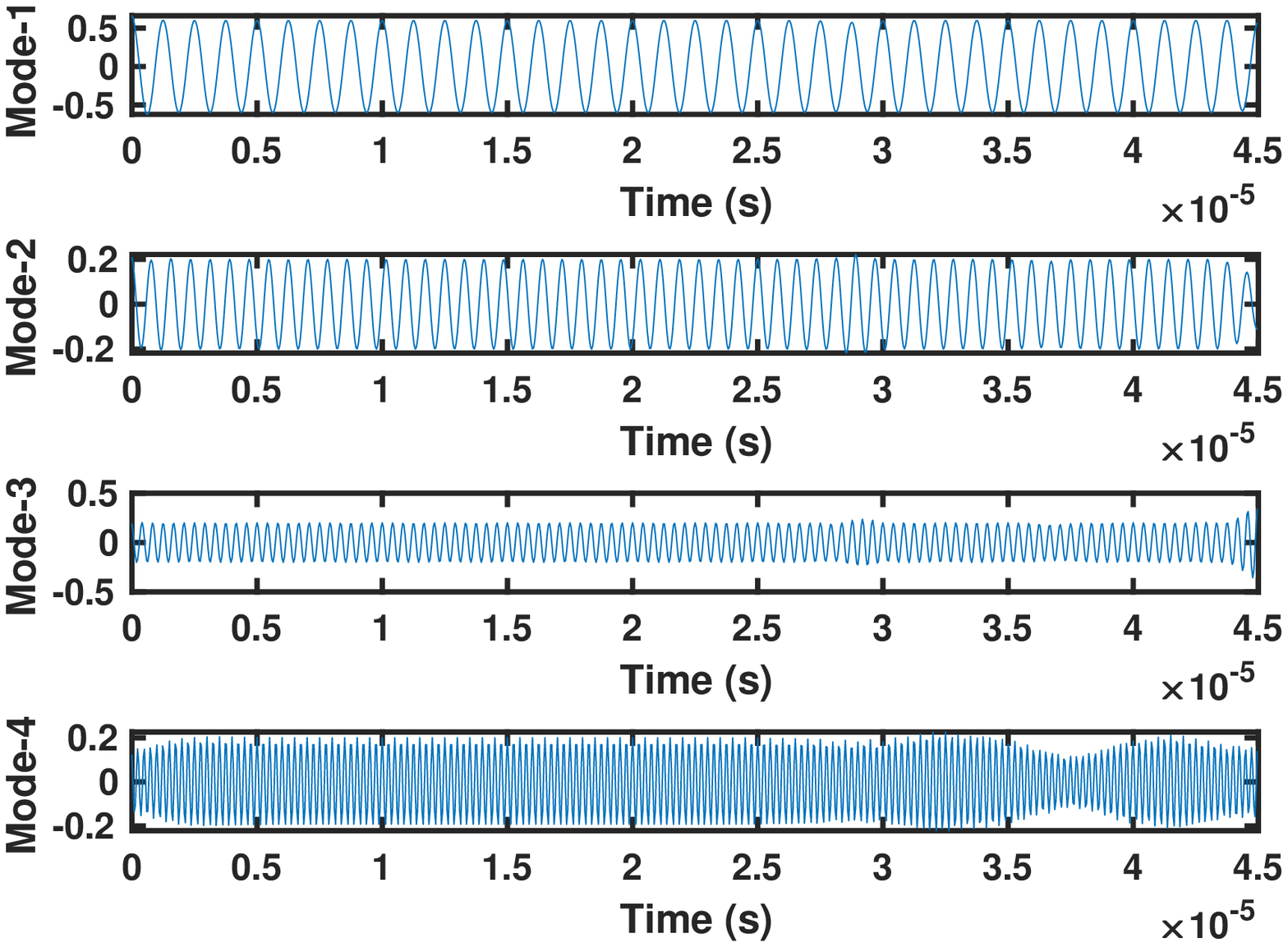}
\end{subfigure}
\begin{subfigure}{.5\textwidth}
  \includegraphics[width=1\linewidth,center]{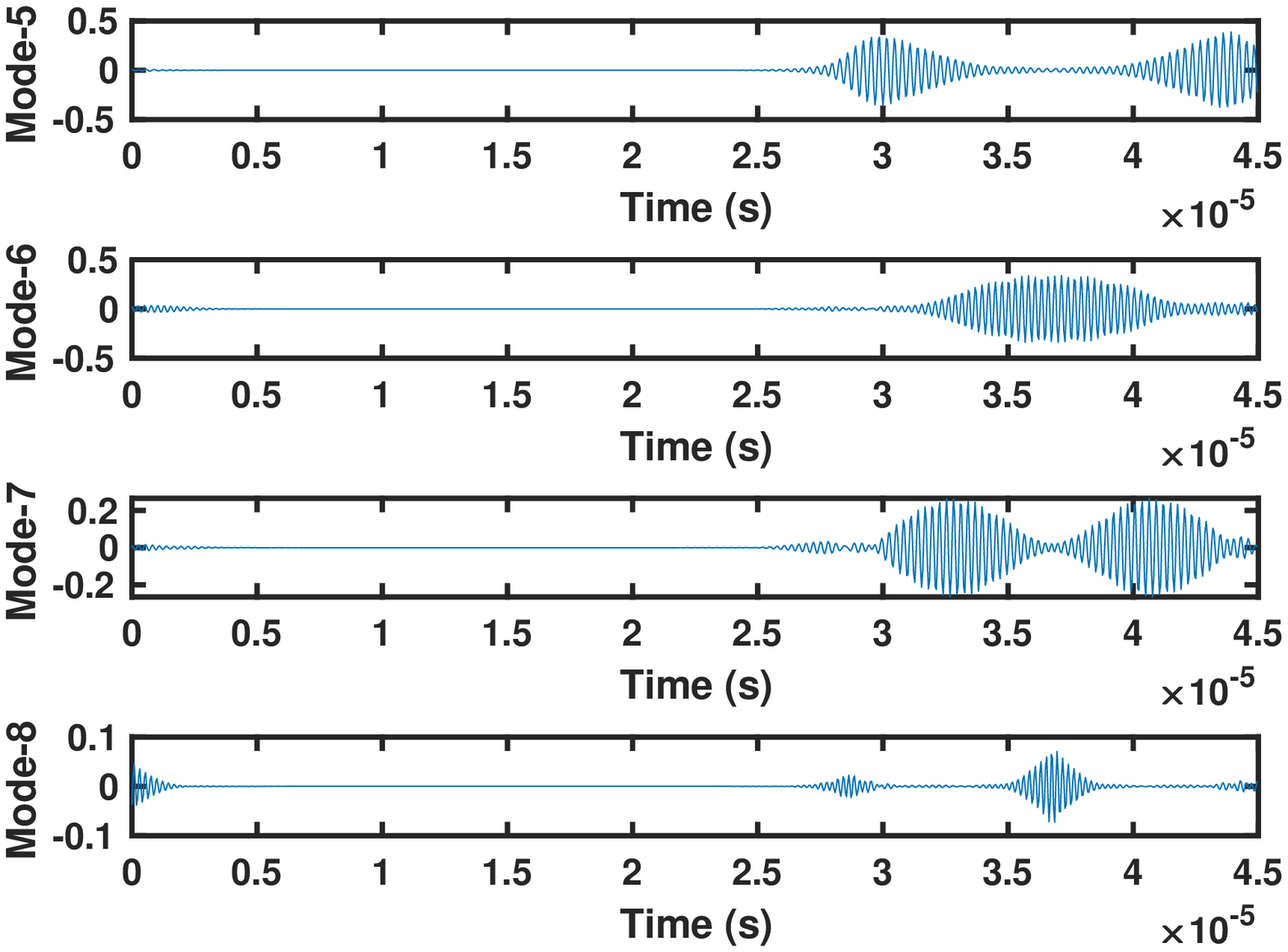}
\end{subfigure}
\caption{VMFs of the received signal containing interference.}
\label{figure:VMD_modes}
\end{figure}

\begin{figure}
  \includegraphics[width=1\linewidth,center]{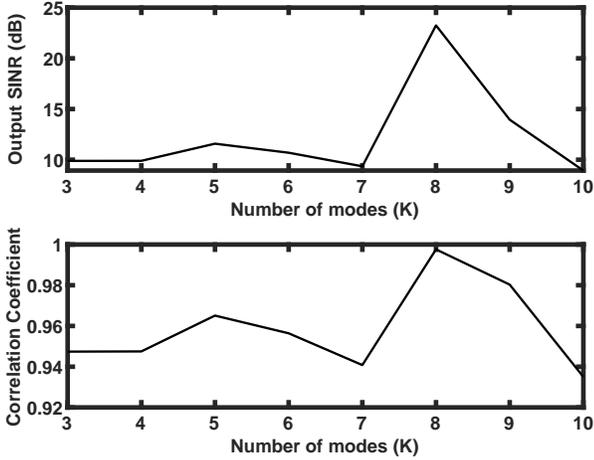}
  \caption{Output SINR and correlation coefficient with respect to number of modes for proposed VMD scheme. 
  }
  \label{fig:fig8}
\end{figure}

\subsubsection{Simulation Results of Proposed VAFER Method 
} 


In the proposed VAFER method, we start with decomposing time domain interference-contaminated received signal by applying proposed VMD algorithm. Fig.~\ref{figure:VMD_modes} shows the resulting $8$ VMFs by selecting $K=8$ as the number of modes. The effectiveness of VMD scheme can be revealed from the figure that the first 4 VMFs showing sinusoid-like variation (mode-1 to mode-4) represent beat frequencies corresponding to 4 targets; while the remaining 4 modes (mode-5 to mode-8) show the interference contaminating the FMCW signal. As can be observed from the most-top plot of Fig.~\ref{figure:VMD_modes} that mode-1 possesses the lowest beat frequency with the largest magnitude, which reflects the largest signal magnitudes in Figs.~\ref{fig:fig1}(c) and \ref{fig:fig1}(d) with yellow color. Furthermore, we intend to verify the feasibility of choosing $K=8$ as the mode number in proposed VAFER scheme. Fig.~\ref{fig:fig8} shows the reconstructed signal strength after interference suppression $SINR_O$ in (\ref{eqn:op_sinr}) and the correlation coefficient $\rho$ in (\ref{rho}) versus the number of mode $K$. It can be seen that the maximal values of both $SINR_O$ and $\rho$ are achieved at $K=8$, which validate the feasible choice on number of modes in our proposed VMD scheme in Fig.~\ref{figure:VMD_modes}.

\begin{figure}
\centering
\begin{subfigure}{.5\textwidth}
  \includegraphics[width=1\linewidth,center]{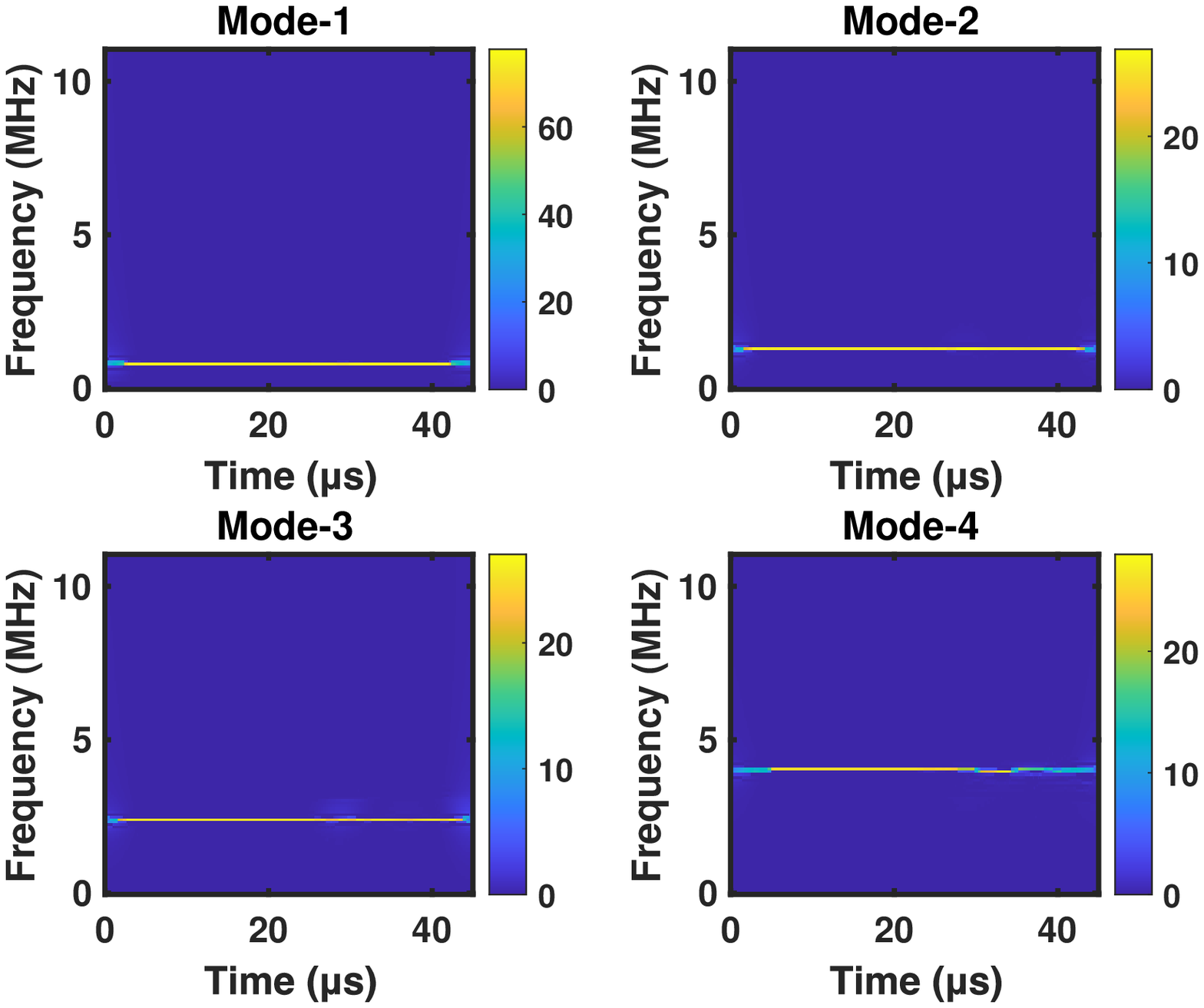}
\end{subfigure}
\begin{subfigure}{.5\textwidth}
  \includegraphics[width=1\linewidth,center]{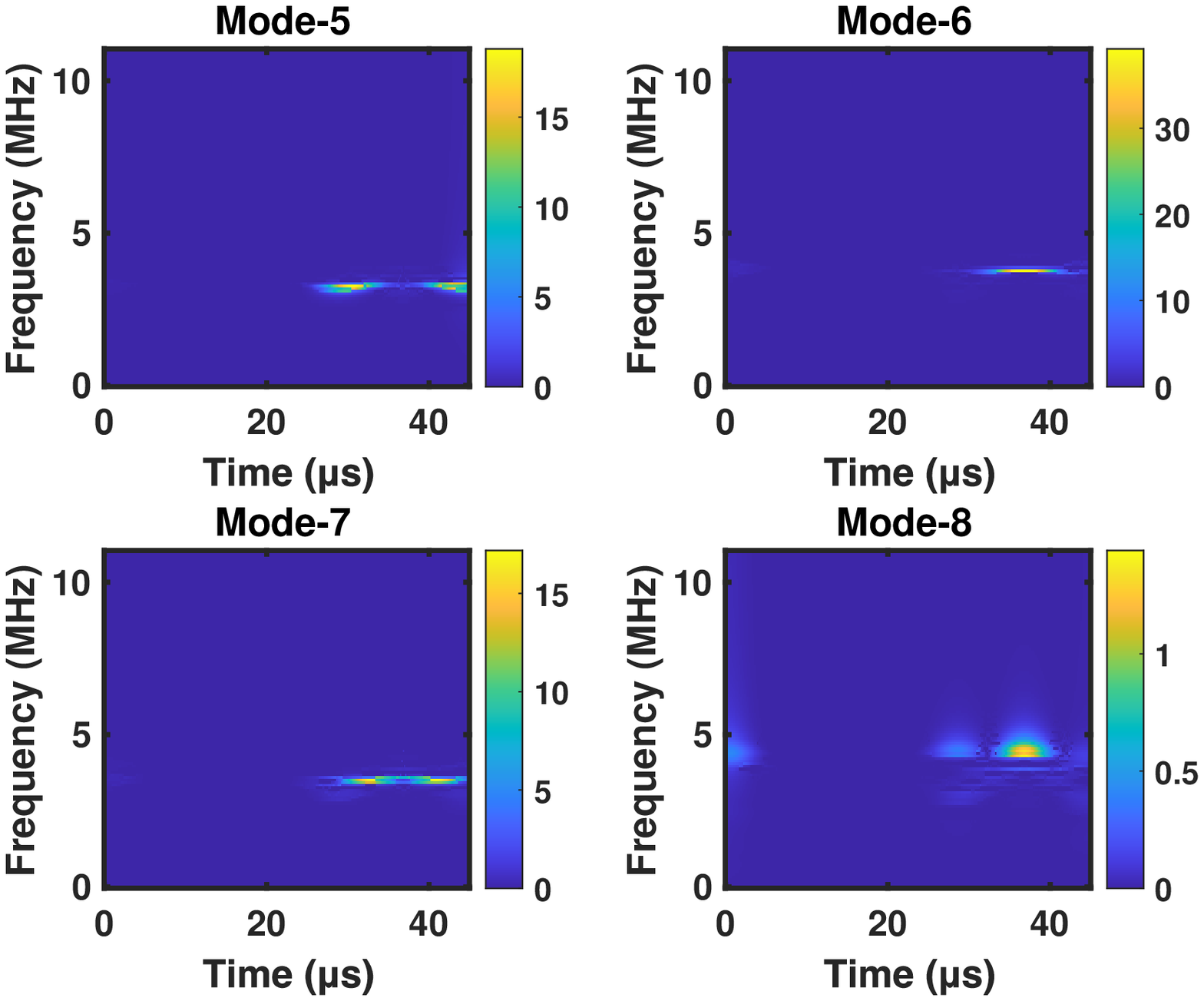}
\end{subfigure}
\caption{Time-frequency spectra obtained by applying FSST to $8$ separated modes after adopting VMD on the interference-corrupted received signal.}\label{figure:fsst}
\end{figure}

Subsequently, the obtained t-f spectra are shown in Fig.~\ref{figure:fsst} by applying FSST on the VMFs. As illustrated in the top four subplots of Fig.~\ref{figure:fsst} representing the t-f spectra of first $4$ VMFs in Fig. \ref{figure:VMD_modes}, each subplot possesses a straight horizontal line parallel to the time axis. The reason is that our proposed FSST sparsely represents VMFs that contain information about target reflections as a single-tone beat frequency in the t-f representation. On the other hand, the bottom $4$ subplots show the t-f spectra of the remaining $4$ VMFs, which reveal high-power noises indicating interference that are separated from the original interference-corrupted signal by adopting the proposed VMD algorithm.

\begin{figure}
    \centering
    \includegraphics[width=1\linewidth,center]{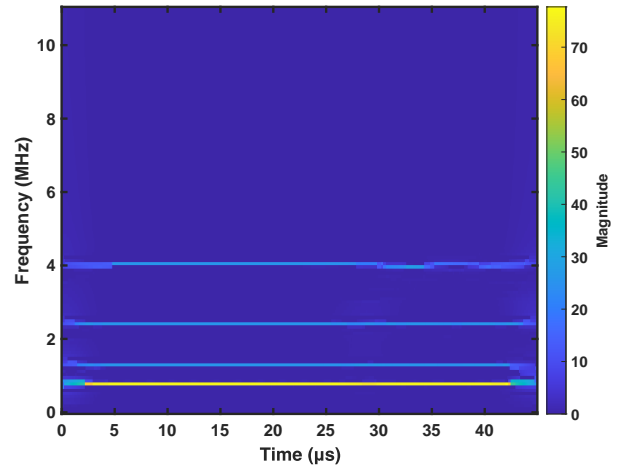}
  \caption{Reconstructed interference suppressed time-frequency spectrum using proposed VAFER method.
  }
    \label{fig:VAFER}
\end{figure}

\begin{figure}[!ht]
\centering
\begin{subfigure}{.49\textwidth}
  \includegraphics[width=1\linewidth,center]{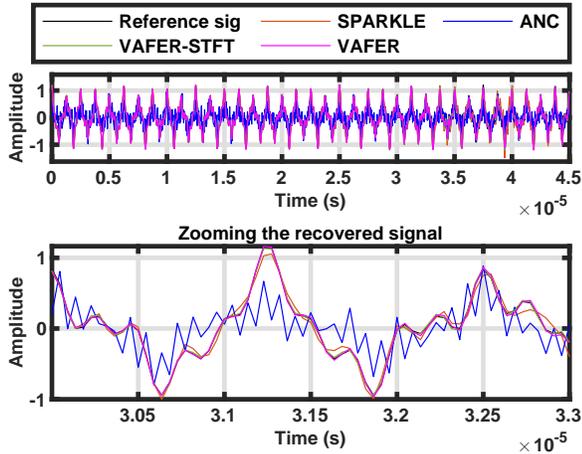}
  \caption{Time domain reconstructed signal 
  }
\end{subfigure}
\begin{subfigure}{.49\textwidth}
  \includegraphics[width=1\linewidth,center]{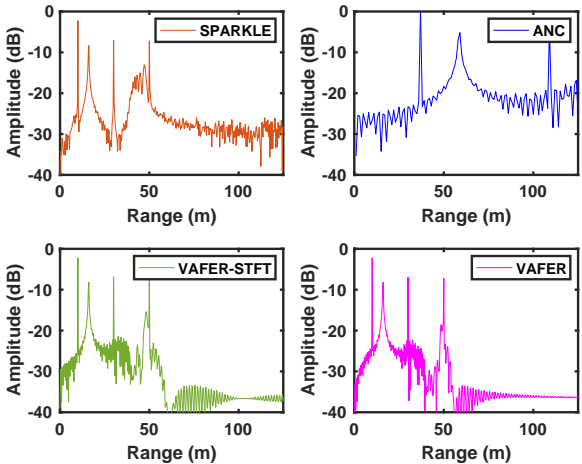}
  \caption{Range DFT of reconstructed signal}
\end{subfigure}
\caption{Performance comparison among SPARKLE~\cite{wang_2022}, ANC~\cite{jin_2019}, STFT-VAFER, and proposed VAFER schemes for reconstructed interference-suppressed (a) time domain and (b) range DFT plots ("Reference sig" denotes interference-free signal as the reference). 
}
\label{fig:fig3}
\end{figure}

Thus, in order to identify the 4 correct modes from the VMD result in Fig.~\ref{figure:fsst}, the energy of VMFs and wiener entropy of time-frequency spectra of VMFs is calculated using \eqref{eq:threshold}, which results in a threshold of $T_{\beta}=0.9681$ for the considered scenario. With the proposed method, interference-contaminated modes are discarded using the computed energy-entropy-based threshold, and interference suppressed signal is reconstructed using the remaining modes. 
Fig. \ref{fig:VAFER} shows the t-f spectrum of reconstructed signal obtained after applying the proposed interference suppression method on the interference-contaminated signal shown in Fig.~\ref{fig:fig1}(d). Fig. \ref{fig:VAFER} demonstrates that interference shown by the box in Fig.~\ref{fig:fig1}(d) is well suppressed by the proposed VAFER method, and the reconstructed t-f spectrum contains only beat frequencies corresponding to the four point targets. Finally, we take inverse FSST to the obtained t-f spectra shown in Fig. \ref{fig:VAFER} to recover the corresponding beat signals, which will be illustrated in Fig.~\ref{fig:fig3} for performance comparison.


\subsubsection{Reconstructed Interference-suppressed Signal Comparison}

For performance comparison with our proposed VAFER scheme, we also perform interference suppression for the simulated interference-corrupted signal in Fig.~\ref{fig:fig1} by implementing the SPARKLE~\cite{wang_2022}, ANC~\cite{jin_2019}, and the STFT variant of VAFER (STFT-VAFER) methods. Figs. \ref{fig:fig3}(a) and \ref{fig:fig3}(b) show the time domain signal and range DFT of the reconstructed signal obtained after applying different interference suppression methods on the interference-corrupted signal. 

Due to interference, it can be observed in Fig.~\ref{fig:fig1}(f) that the noise floor has increased to raise the challenge of detecting the correct target range. The results in Fig.~\ref{fig:fig3}(a) show that time domain reconstructed signal using proposed VAFER method coincides with the trends of the reference signal whereas the results of ANC are having ripples in the time domain reconstructed signal. It can be observed that although time domain reconstructed signal obtained by applying SPARKLE and STFT version of VAFER methods are close to the reference signal, the overall performance of proposed VAFER method is superior as validated in Fig.~\ref{fig:fig3}(a). Moreover, the results in Fig.~\ref{fig:fig3}(b) show that after applying the proposed VAFER method noise floor has been reduced significantly around the target peak at $50$ meters. 
It can be seen that although SPARKLE method reduces noise floor significantly, a total of $5$ peaks were detected for $4$ 
targets; whereas the proposed VAFER scheme detects exact $4$ peaks with a reduction in noise floor as well. On the other hand, the performance of ANC method is poor due to the assumption in \cite{jin_2019} about beat frequency occupying only the positive half of the spectrum, and interference in the negative half of the spectrum is conjugate symmetric of the positive spectrum. Also, benefiting from the property of FSST to focus more on sinusoid-like beat signals \cite{thakur_2011}, the performance of VAFER is superior to the STFT-VAFER as shown in both Figs.~\ref{fig:fig3}(a) and \ref{fig:fig3}(b).

Furthermore, We evaluate $SINR_O$ in (\ref{eqn:op_sinr}) and correlation coefficient $\rho$ in (\ref{rho})
 of the reconstructed signals using the same setting in Subsection \ref{sec:simsetup}. 
The corresponding results are shown in Table ~\ref{table:2}. Compared to the SPARKLE, ANC, and VAFER-STFT methods, it can be clearly observed that our proposed VAFER algorithm results in the highest $SINR_O=23.2504$ dB and $\rho = 0.9976$ by suppressing the interference.
The VAFER scheme significantly improves the reconstructed signal's SINR by $14.07$ dB based on the original interference-corrupted received signal of $SINR_I = 9.1814$ dB. These results validate that our proposed VAFER method can suppress interference and improve signal quality compared to existing methods.

\begin{table}[ht]
\centering
\caption{\\ SINR comparison of reconstructed signals}
\begin{tabular}[t]{lcc}
\hline
Method & $SINR_o$ (dB) & Correlation Coefficient $(\rho)$ \\
\hline
SPARKLE & 14.5503 & 0.9828 \\
ANC & -7.0505 & 0.0012 \\
VAFER-STFT & 19.3525 & 0.9942\\
VAFER &  23.2504 & 0.9976\\
\hline
\end{tabular}
\label{table:2}
\end{table}

\begin{figure}[!ht]
\begin{subfigure}{.48\textwidth}
  \includegraphics[width=1\linewidth,center]{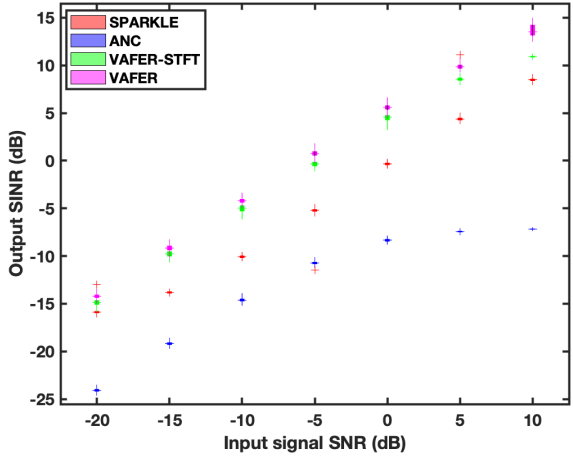}
  \caption{}
\end{subfigure}
\begin{subfigure}{.48\textwidth}
  \includegraphics[width=1\linewidth,center]{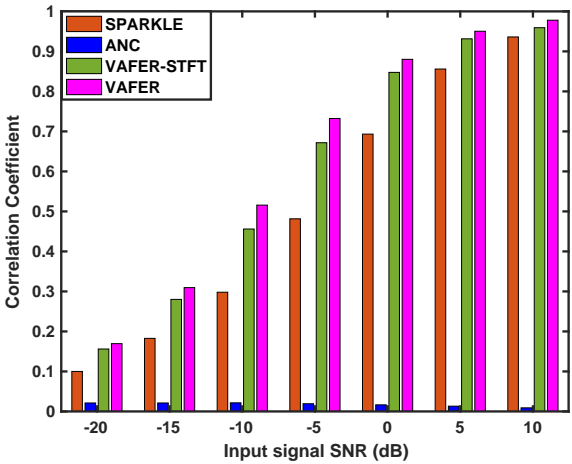}
  \caption{}
\end{subfigure}
\caption{Quantitative analysis of the interference suppression performance of the SPARKLE, ANC, VAFER-STFT, and our proposed VAFER method at varying SNRs of the input signals from $-20$ to $10$ dB. (a) and (b) respectively show the corresponding variations of output SINRs $SINR_O$ and the correlation coefficients $\rho$ of interference-suppressed beat signals.}
\label{fig:fig_box}
\end{figure}

\begin{figure}[!ht]
\centering
\begin{subfigure}{.49\textwidth}
  \includegraphics[width=1\linewidth,center]{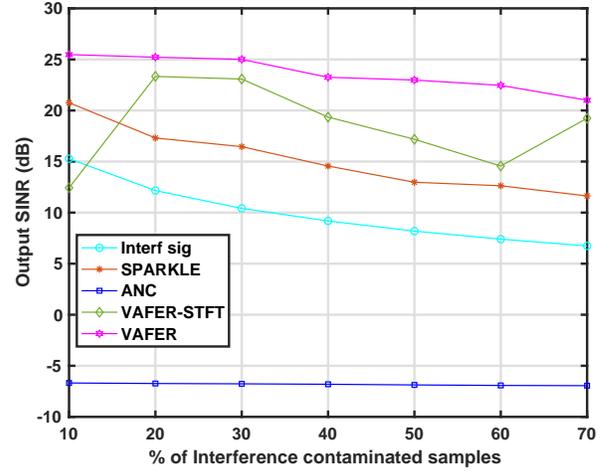}
  \caption{}
\end{subfigure}
\begin{subfigure}{.49\textwidth}
  \includegraphics[width=1\linewidth,center]{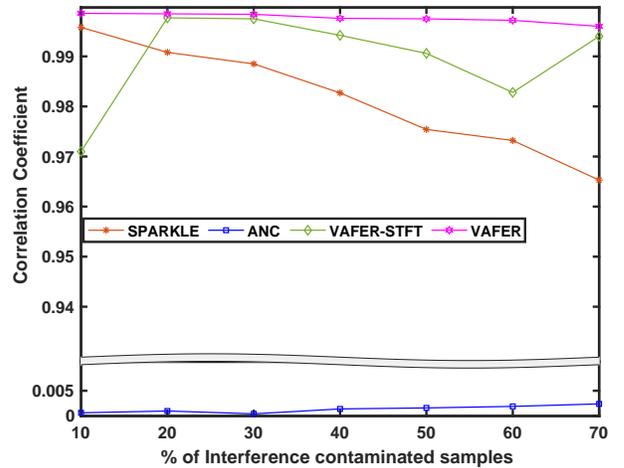}
  \caption{}
\end{subfigure}
\caption{
Quantitative analysis of the interference suppression performance of the SPARKLE, ANC, VAFER-STFT, and our proposed VAFER method at different percentages of interference duration ("Interf sig" denotes interference-contaminated signal). (a) and (b) respectively show the corresponding variations of output SINRs $SINR_O$ and the correlation coefficients $\rho$ of interference-suppressed beat signals.}
\label{fig:fig7}
\end{figure}

\subsubsection{Effect of SNR}

We examine the effect of SNR on the interference suppression performance by varying the added AWGN noise levels of the proposed VAFER method. We consider $J = 3$ targets located at $R=[20, 40, 70]$ meters and performed $300$ Monte Carlo runs at each noise level of SNR varying from $-20$ to $10$ dB. The radar is simulated following parameters of Table ~\ref{table:1}. The results corresponding to the output SINR $SINR_O$ in (\ref{eqn:op_sinr}) of reconstructed signal for stated Monte Carlo runs are illustrated as box plot in Fig.~\ref{fig:fig_box}(a), where the $25^{th}$ and $75^{th}$ percentiles of results are indicated by the bottom and top of each box. On the other hand, the bar chart in Fig.~\ref{fig:fig_box}(b) shows the comparison in correlation coefficient $\rho$ in (\ref{rho}) as SNR is varied from $-20$ to $10$ dB. We can intuitively observe from both figures that both output SINR and correlation coefficient increase with the SNR of input signal for all schemes except ANC method. Compared with existing three schemes, our proposed VAFER method can achieves the highest output SINR and correlation coefficient under different SNR values of input signal.

\subsubsection{Performance Evaluation for Varying Interference Duration}
We evaluate the performance of VAFER algorithm by varying the percentage of interference-contaminated samples in the simulated signal using the same setting in Subsection \ref{sec:simsetup}. We vary the percentage of interference-contaminated samples from $10\%$ to $70\%$ at constant $SINR_I$ of $9.1814$ dB and compute the corresponding $SINR_O$ and $\rho$. The interference-corrupted beat signal denoted by the legend "Interf sig" in Fig.~\ref{fig:fig7}(a) is simulated by varying the chirp duration of the interference signal. Performance comparison are illustrated in Figs.~\ref{fig:fig7}(a) and \ref{fig:fig7}(b) where we show output SINR in (\ref{eqn:op_sinr}) and $\rho$ in (\ref{rho}) versus $\%$ of interference-contaminated received signal, respectively. It can be observed in Fig.~\ref{fig:fig7}(a) 
that $SINR_O$ for ANC method is almost constant and is the lowest compared to the other schemes. On the other hand, Fig.~\ref{fig:fig7}(b) shows that the correlation coefficients for SPARKLE, VAFER-STFT, and proposed VAFER method are almost similar; whereas the ANC scheme results in the least $\rho$ value. It can be seen in both plots that our proposed VAFER scheme achieves the highest $SINR_O$ and $\rho$ than SPARKLE, ANC, and VAFER-STFT methods. 

Furthermore, as shown in Fig.~\ref{fig:fig7}(a) and ~\ref{fig:fig7}(b), there are declined values of $SINR_O$ and $\rho$ in the VAFER-STFT method under both $10 \%$ and $60 \%$ of signal samples contaminating with interference. The main reason is that the energy-entropy threshold fails to discard some of the modes containing interference in VAFER-STFT leading to lowered $SINR_O$ and $\rho$ values. Here we also observe that as the number of interference-contaminated samples in the signal increases, $SINR_O$ and $\rho$ decrease slightly due to some residuals of interference in the VMD modes corresponding to targets in the proposed VAFER method. On the other hand, the assumption of interference in the negative half of spectrum being conjugate symmetric of the positive spectrum for ANC method leads to a slightly increase in the $\rho$ value.

\subsection{Experimental Results}

The experimental radar data and radar parameters for this section have been taken from the recent work carried out by authors Feng Jin and Siyang Cao in~\cite{jin_2019}.
In their experimental setup, interference-contaminated data is obtained by considering a car traveling with the speed of $15$ m/s towards the victim FMCW radar operating at $77$ GHz while being affected by an interfering FMCW radar operating at the same $77$ GHz at a distance of $2$ meters from the victim radar. The experimental parameters are available in Table ~\ref{table:3}. 

\begin{figure}[!ht]
\begin{subfigure}{.46\textwidth}
\includegraphics[width=01\linewidth,center]{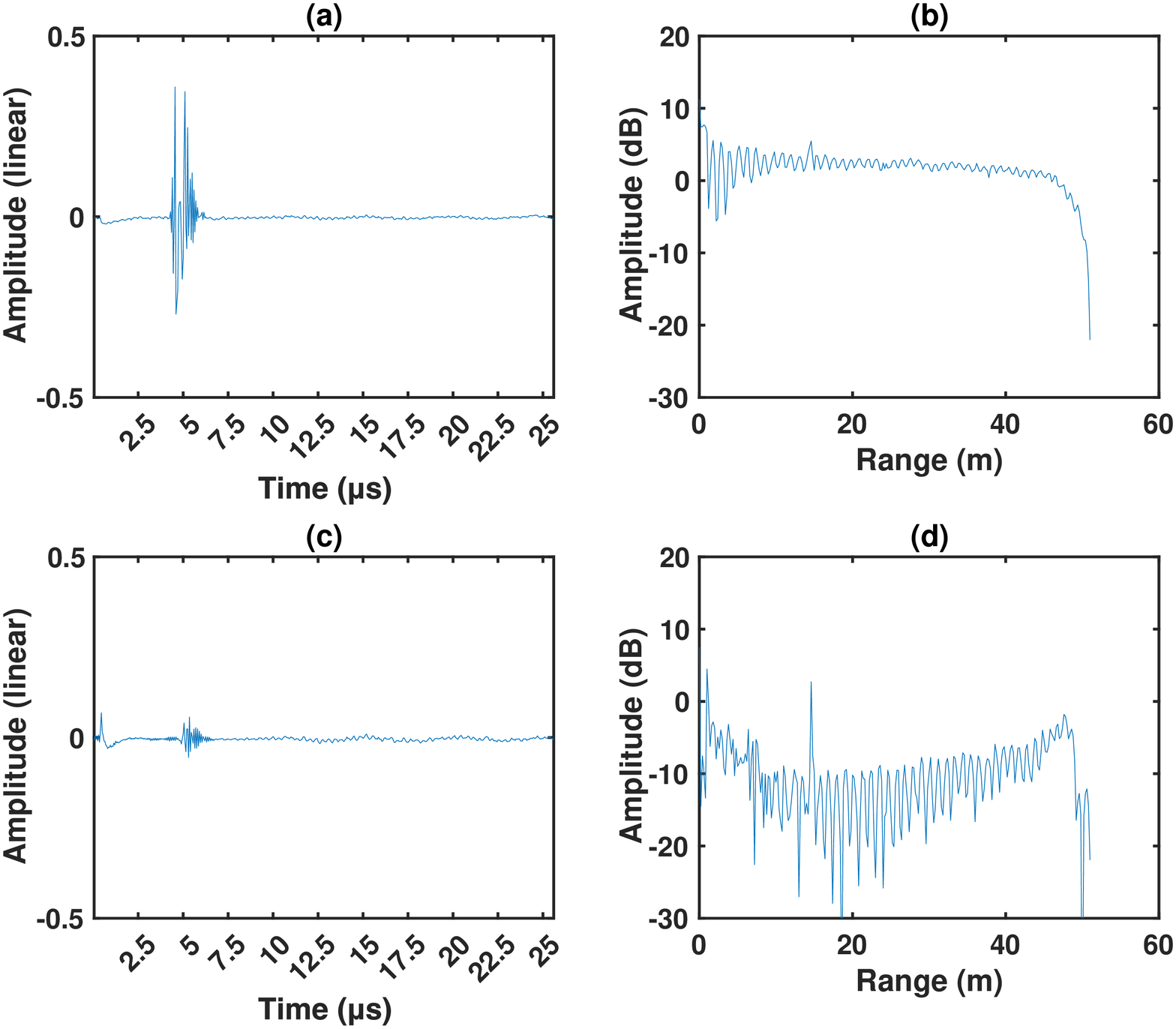}
  \label{fig:sfig6}
\end{subfigure}%
\\
\begin{subfigure}{.46\textwidth}
  \includegraphics[width=01\linewidth,center]{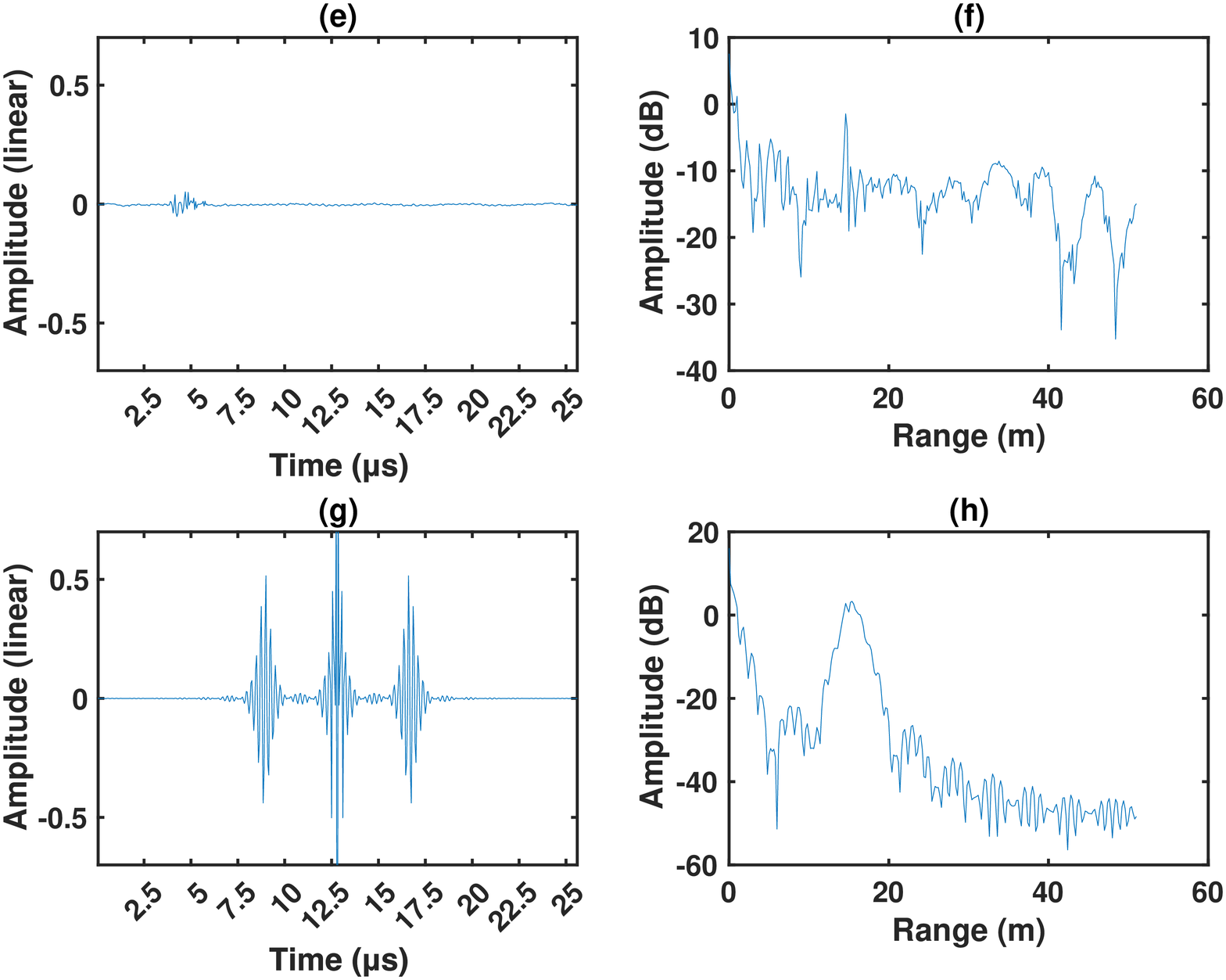}
  \label{fig:sfig_6}
\end{subfigure}%
\\
\begin{subfigure}{0.46\textwidth}
\includegraphics[width=1\linewidth,center]{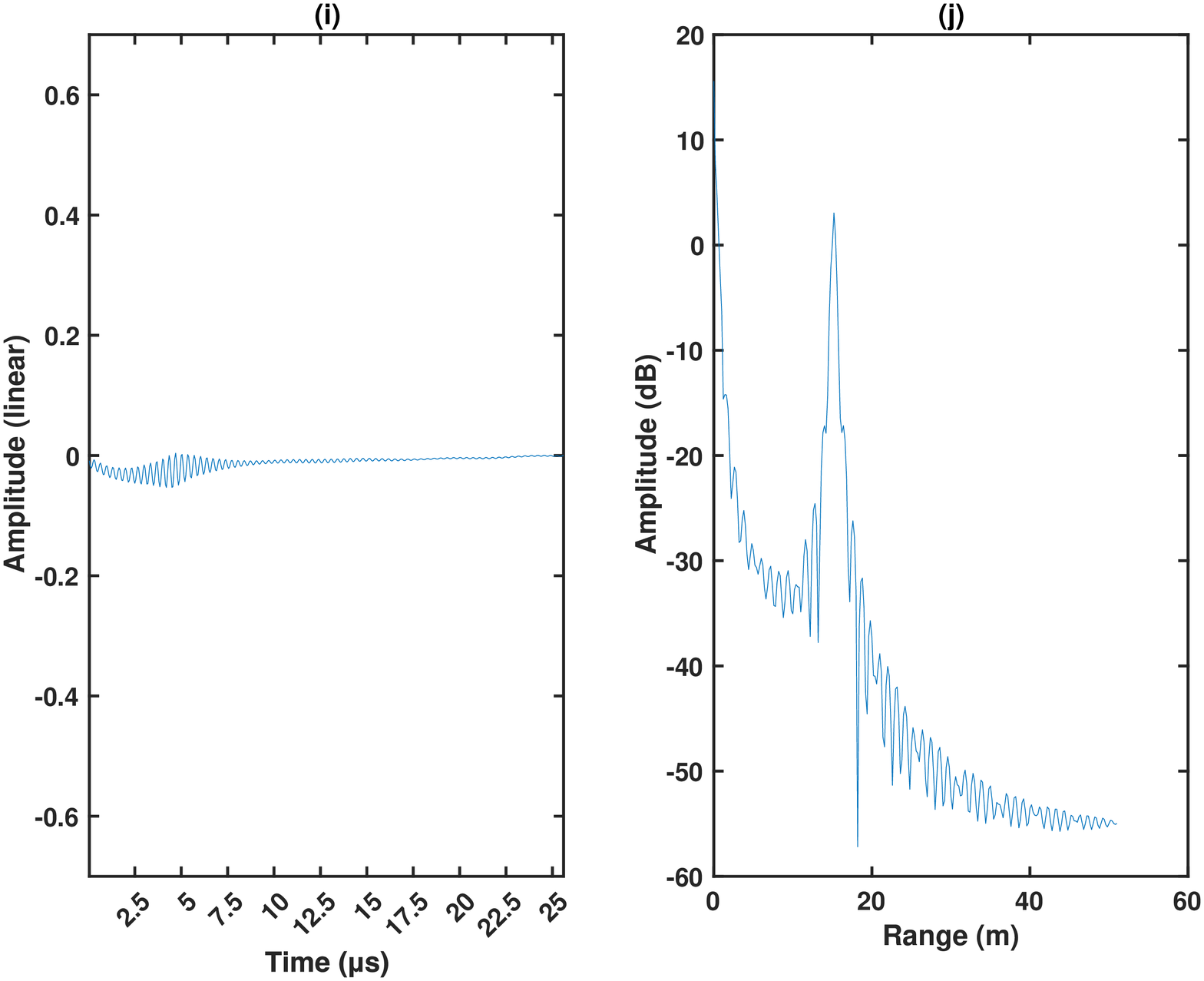}
  \label{fig:s_fig6}
\end{subfigure}%
\caption{Performance comparison via experimental results showing time domain representation (left-column subplots) and range profile (right-column subplots). (a) and (b): interference-contaminated signals; (c) and (d): ANC; (e) and (f): SPARKLE,\;  (g) and (h): VAFER-STFT; (i) and (j): proposed VAFER method.}
\label{fig:exp_res}
\end{figure}

\begin{table}
    \centering
    \caption{\\ Experimental Parameters 
    }
\begin{tabular}{ c c c  } 
\hline
Radar Type & Parameter & Value \\
\hline
Victim Radar & Carrier Frequency $f_o$ & 77 GHz \\ 
& Bandwidth $B$ & 750 MHz \\ 
& Chirp Duration $T$ & 29.56 $\mu$sec  \\ 
& Sampling Frequency $f_s$ & 20 Msps \\
\hline
Interfering Radar & Carrier Frequency $f_m$ & 77 GHz \\ 
& Bandwidth $B_{int}$ & 682 MHz \\ 
& Chirp Duration $T_{int}$ & 72.31 $\mu$sec \\ 
& Sampling Frequency $f_{s,int}$ & 15 Msps \\
\hline
\end{tabular}

    \label{table:3}
\end{table}
For this study, we have compared our proposed VAFER scheme with ANC, SPARKLE, and VAFER-STFT methods. Fig.~\ref{fig:exp_res} shows the corresponding results of time domain reconstructed interference-suppressed signals shown as the linear amplitudes of received signal versus time (left-column subplots) and the range profiles obtained by performing DFT on fast time shown as DFT amplitude in dB versus range in meters (right-column subplots).
The interference-contaminated time domain signal and range profile are shown in Fig.~\ref{fig:exp_res}(a) and ~\ref{fig:exp_res}(b), respectively, where the interference is visible as high fluctuating peaks from $3$ to $6$ $\mu$sec in Fig.~\ref{fig:exp_res}(a). Notice that the target is located at $14.8$ meters which is not clearly revealed in Fig.~\ref{fig:exp_res}(b). The results of ANC scheme are shown in Figs. \ref{fig:exp_res}(c) and \ref{fig:exp_res}(d) which still indicate the presence of some interference, and a number of additional peaks are detected in the range profile showing false or ghost targets. The interference suppression results for SPARKLE algorithm are demonstrated in Figs. \ref{fig:exp_res}(e) and \ref{fig:exp_res}(f). Results show that although the method performs well in suppressing the interference in time domain, still the range profile shows the presence of false targets. In the range profiles obtained by ANC and SPARKLE in Figs. \ref{fig:exp_res}(d) and \ref{fig:exp_res}(f) respectively, it can be observed that both schemes also detect the presence of interfering radar at $2$ meter as an additional target which limits their performance and reducing the output SINR values. Moreover, the results in Figs.~\ref{fig:exp_res}(g) and \ref{fig:exp_res}(h) show that the VAFER-STFT method is ineffective for suppressing interference resulting in excessive time domain amplitudes. Figs.~\ref{fig:exp_res}(i) and \ref{fig:exp_res}(j) demonstrate the ability of our proposed VAFER method on interference suppression, where Fig.~\ref{fig:exp_res}(i) shows that the peak fluctuations caused by interfering radar are significantly suppressed and only a single peak is obtained corresponding to the actual target located at $14.8$ meters as illustrated in Fig.~\ref{fig:exp_res}(j).

For quantitative performance evaluation, we compare the SINR values computed by following the formulation used in ~\cite{jin_2019} on their experimental data after applying the compared schemes. Here we compute SINR based on their predefined empirical parameters such as guard cells and reference cells defined around the point target. Notice that the SINR computation using \eqref{eqn:op_sinr} requires a reference signal and it is computed for the entire signal which is not only around the target peak. Note that due to unavailability of raw interference-free data, the computation of correlation coefficient is not feasible for this experimental setup. As a result, the SINR of interference-contaminated data is $2.94$ dB; whereas that of  interference-suppressed signal after applying the ANC method is $4.97$ dB with an improvement of $2.03$ dB. The output SINR of SPARKLE methods is $9.62$ dB with an improvement of $6.67$ dB and the VAFER-STFT method improves SINR by $6.66$ dB resulting in an output SINR of $9.61$ dB. The proposed VAFER method improves the SINR by $9.87$ dB resulting in an output SINR of $12.81$ dB. The merits of VAFER scheme can therefore be observed.

\section{Conclusion}

In this paper, we propose an interference suppression method, VAFER, based on VMD and FSST for FMCW radar. After a time-frequency analysis of VMD modes, we proposed to use an energy entropy-based threshold to eliminate interference-contaminated modes. The effectiveness of our proposed VAFER scheme has been verified by comparing it with two recently proposed state-of-the-art methods, ANC and SPAARKLE, and has been found to significantly improve the output SINR with higher correlation coefficient for both simulated and experimental data. Also, the effect of varying time duration of interference on output SINR and correlation coefficient has been studied. It is inferred that our proposed VAFER method can significantly suppress the interference without any observable degradation in performance. Our proposed approach is expected to pave way for a new paradigm in mutual interference suppression for FMCW radars. 

\bibliography{refs}

\begin{thebibliography}{10}

\bibitem{bilik_2019}
I.~Bilik, O.~Longman, S.~Villeval, and J.~Tabrikian, ``The rise of radar for
  autonomous vehicles: Signal processing solutions and future research
  directions,'' {\em IEEE Signal Process. Mag.}, vol.~36, no.~5, pp.~20--31,
  2019.

\bibitem{Luo_2013}
T.-N. Luo, C.-H.~E. Wu, and Y.-J.~E. Chen, ``A 77-\uppercase{GH}z \uppercase
  {CMOS} automotive radar transceiver with anti-interference function,'' {\em
  IEEE Trans. Circuits Syst. I: Regul. Pap.}, vol.~60, no.~12, pp.~3247--3255,
  2013.

\bibitem{alland_2019}
S.~Alland, W.~Stark, M.~Ali, and M.~Hegde, ``Interference in automotive radar
  systems: Characteristics, mitigation techniques, and current and future
  research,'' {\em IEEE Signal Process. Mag.}, vol.~36, no.~5, pp.~45--59,
  2019.

\bibitem{Bourdoux_2017}
A.~Bourdoux, K.~Parashar, and M.~Bauduin, ``Phenomenology of mutual
  interference of \uppercase{FMCW} and \uppercase {PMCW} automotive radars,''
  in {\em 2017 IEEE Radar Conference (RadarConf)}, pp.~1709--1714, 2017.

\bibitem{uysal_2020}
F.~Uysal, ``Phase-coded \uppercase {FMCW} automotive radar: System design and
  interference mitigation,'' {\em IEEE Trans. Veh. Technol.}, vol.~69, no.~1,
  pp.~270--281, 2020.

\bibitem{khoury_2016}
J.~Khoury, R.~Ramanathan, D.~McCloskey, R.~Smith, and T.~Campbell,
  ``Radar\uppercase{MAC}: Mitigating radar interference in self-driving cars,''
  in {\em 2016 13th Annual IEEE International Conference on Sensing,
  Communication, and Networking (SECON)}, pp.~1--9, 2016.

\bibitem{Aydogdu_2021}
C.~Aydogdu, M.~F. Keskin, N.~Garcia, H.~Wymeersch, and D.~W. Bliss,
  ``Rad\uppercase{C}hat: Spectrum sharing for automotive radar interference
  mitigation,'' {\em IEEE Trans. Intell. Transp. Syst.}, vol.~22, no.~1,
  pp.~416--429, 2021.

\bibitem{Bechter_2016}
J.~Bechter, C.~Sippel, and C.~Waldschmidt, ``Bats-inspired frequency hopping
  for mitigation of interference between automotive radars,'' in {\em 2016 IEEE
  MTT-S International Conference on Microwaves for Intelligent Mobility
  (ICMIM)}, pp.~1--4, 2016.

\bibitem{xu_2018}
Z.~Xu and Q.~Shi, ``Interference mitigation for automotive radar using
  orthogonal noise waveforms,'' {\em IEEE Geosci. Remote. Sens. Lett.},
  vol.~15, no.~1, pp.~137--141, 2018.

\bibitem{Irazoqui_2019}
R.~W. Irazoqui and C.~J. Fulton, ``Spatial interference nulling before
  \uppercase{RF} frontend for fully digital phased arrays,'' {\em IEEE Access},
  vol.~7, pp.~151261--151272, 2019.

\bibitem{Hu_2019}
X.~Hu, Y.~Li, M.~Lu, Y.~Wang, and X.~Yang, ``A multi-carrier-frequency
  random-transmission chirp sequence for \uppercase {TDM MIMO} automotive
  radar,'' {\em IEEE Trans. Veh. Technol.}, vol.~68, no.~4, pp.~3672--3685,
  2019.

\bibitem{Kunert_2012}
M.~Kunert, ``The \uppercase{EU} project \uppercase{MOSARIM}: A general overview
  of project objectives and conducted work,'' in {\em 2012 9th European Radar
  Conference}, pp.~1--5, 2012.

\bibitem{jin_2019}
F.~Jin and S.~Cao, ``Automotive radar interference mitigation using adaptive
  noise canceller,'' {\em IEEE Trans. Veh. Technol.}, vol.~68, no.~4,
  pp.~3747--3754, 2019.

\bibitem{chen_2018}
Z.~Chen, F.~Xie, C.~Zhao, and C.~He, ``Radio frequency interference
  cancellation in high-frequency surface wave radar using orthogonal projection
  filtering,'' {\em IEEE Geosci. Remote Sens. Lett.}, vol.~15, no.~9,
  pp.~1322--1326, 2018.

\bibitem{wang_2021}
J.~Wang, M.~Ding, and A.~Yarovoy, ``Matrix-pencil approach-based interference
  mitigation for \uppercase{FMCW} radar systems,'' {\em IEEE Trans. Microw.
  Theory Techn.}, vol.~69, no.~11, pp.~5099--5115, 2021.

\bibitem{lee_2021}
S.~Lee, J.-Y. Lee, and S.-C. Kim, ``Mutual interference suppression using
  wavelet denoising in automotive \uppercase{FMCW} radar systems,'' {\em IEEE
  Trans. Intell. Transp. Syst}, vol.~22, no.~2, pp.~887--897, 2021.

\bibitem{xu_2021}
Z.~Xu and M.~Yuan, ``An interference mitigation technique for automotive
  millimeter wave radars in the tunable \uppercase{Q} -factor wavelet transform
  domain,'' {\em IEEE Trans. Microw. Theory Tech.}, vol.~69, no.~12,
  pp.~5270--5283, 2021.

\bibitem{Neemat_2019}
S.~Neemat, O.~Krasnov, and A.~Yarovoy, ``An interference mitigation technique
  for \uppercase{FMCW} radar using beat-frequencies interpolation in the
  \uppercase{STFT} domain,'' {\em IEEE Trans. Microw. Theory Tech.}, vol.~67,
  no.~3, pp.~1207--1220, 2019.

\bibitem{Uysal_2019}
F.~Uysal, ``Synchronous and asynchronous radar interference mitigation,'' {\em
  IEEE Access}, vol.~7, pp.~5846--5852, 2019.

\bibitem{Liu_2020}
Z.~Liu, W.~Lu, J.~Wu, S.~Yang, and G.~Li, ``A \uppercase{PELT-KCN} algorithm
  for \uppercase {FMCW} radar interference suppression based on signal
  reconstruction,'' {\em IEEE Access}, vol.~8, pp.~45108--45118, 2020.

\bibitem{Mun_2020}
J.~Mun, S.~Ha, and J.~Lee, ``Automotive radar signal interference mitigation
  using \uppercase {RNN} with self attention,'' in {\em 2020 IEEE International
  Conference on Acoustics, Speech and Signal Processing (ICASSP)},
  pp.~3802--3806, 2020.

\bibitem{wang_2022_2}
J.~Wang, ``\uppercase{CFAR}-based interference mitigation for \uppercase {FMCW}
  automotive radar systems,'' {\em IEEE Trans. Intell. Transp. Syst.}, vol.~23,
  no.~8, pp.~12229--12238, 2022.

\bibitem{wang_2022}
J.~Wang, M.~Ding, and A.~Yarovoy, ``Interference mitigation for
  \uppercase{FMCW} radar with sparse and low-rank \uppercase{H}ankel matrix
  decomposition,'' {\em IEEE Trans. Signal Process.}, vol.~70, pp.~822--834,
  2022.

\bibitem{dragomi_2014}
K.~Dragomiretskiy and D.~Zosso, ``Variational mode decomposition,'' {\em IEEE
  Trans. Signal Process.}, vol.~62, no.~3, pp.~531--544, 2014.

\bibitem{thakur_2011}
G.~Thakur and H.-T. Wu, ``Synchrosqueezing-based recovery of instantaneous
  frequency from nonuniform samples,'' {\em SIAM J. Math. Anal.}, vol.~43,
  no.~5, pp.~2078--2095, 2011.

\bibitem{auger_2013}
F.~Auger, P.~Flandrin, Y.-T. Lin, S.~McLaughlin, S.~Meignen, T.~Oberlin, and
  H.-T. Wu, ``Time-frequency reassignment and synchrosqueezing: An overview,''
  {\em IEEE Signal Process. Mag.}, vol.~30, no.~6, pp.~32--41, 2013.

\bibitem{oberlin_2014}
T.~Oberlin, S.~Meignen, and V.~Perrier, ``The \uppercase{F}ourier-based
  synchrosqueezing transform,'' in {\em 2014 IEEE International Conference on
  Acoustics, Speech and Signal Processing (ICASSP)}, pp.~315--319, 2014.

\bibitem{Dubnov_2004}
S.~Dubnov, ``Generalization of spectral flatness measure for
  non-\uppercase{G}aussian linear processes,'' {\em IEEE Signal Process.
  Lett.}, vol.~11, no.~8, pp.~698--701, 2004.

\end{thebibliography}

\end{document}